\documentclass[11pt,a4paper]{article}
\usepackage{jheppub}

\def\beq{\begin{equation}}
\def\eeq{\end{equation}}

\def\beq{\begin{equation}}
\def\eeq{\end{equation}}
\def\bea{\begin{eqnarray}}
\def\eea{\end{eqnarray}}

\def\eq#1{{Eq.~(\ref{#1})}}
\def\fig#1{{Fig.~\ref{#1}}}

\parskip=\itemsep               
\def\eq#1{{Eq.~(\ref{#1})}}

\def\fig#1{{Fig.~\ref{#1}}}
\title{  Reggeon Field Theory and Self Duality: Making Ends Meet.}
\author[a]{Alex Kovner }
\author[b,c]{\!\!, Eugene Levin}
\author[a]{\!\!, Ming Li }
\author[d]{and Michael Lublinsky}
\affiliation[a]{Physics Department, University of Connecticut, 2152 Hillside Road, Storrs, CT 06269, USA}
\affiliation[b]{Departemento de F\'isica, Universidad T\'ecnica Federico Santa Mar\'ia, and Centro Cient\'ifico-\\
Tecnol\'ogico de Valpara\'iso, Avda. Espana 1680, Casilla 110-V, Valpara\'iso, Chile}
\affiliation[c]{Department of Particle Physics, Tel Aviv University, Tel Aviv 69978, Israel}
\affiliation[d]{Physics Department, Ben-Gurion University of the Negev, Beer Sheva 84105, Israel}

\abstract{Motivated by the question of unitarity of Reggeon Field Theory, we use the effective field theory philosophy to find possible Reggeon Field Theory Hamiltonians $H_{RFT}$. We require that $H_{RFT}$  is self dual, reproduce all known limits (dilute-dense and dilute-dilute) and exhibits all the symmetries of the JIMWLK Hamiltonian. We find a family of Hamiltonians which satisfy all the above requirements. One of these is identical in form to the so called "diamond action" discussed in \cite{diamond,Balitsky05}. However we show by explicit calculation that the so called "diamond condition" is not satisfied beyond leading perturbative order. }

\begin{document}
\maketitle

\pagestyle{empty}
\newpage

\mbox{}

\pagestyle{plain}

\setcounter{page}{1}
\author{}

\abstract{ }
\dedicated{}
\preprint{}


\section{Introduction.}

 Reggeon Field Theory (RFT) of Quantum Chromodynamics (QCD) is a putative effective theory that is meant to describe scattering at asymptotically high energies. Development of this theory  during the last three decades lead to understanding of many features of high energy scattering as well as phenomenological applications to HERA, RHIC and LHC data. Nevertheless, much work notwithstanding the theoretical framework of RFT is incomplete.

The basic pre-QCD ideas of RFT go back to Gribov \cite{gribov}, who considered a very general picture and properties of high energy exchanges in a local field theory. These ideas have been adopted to QCD and furhter developed over the years in many works \cite{BFKL,glr,MUPA,MUDI,LIREV,LipatovFT, bartels,BKP, KLLL,KLL,mv,Salam, KOLE,BRN,braun,BK,AKLL,AKLL1}.  Direct derivation of some elements of RFT from QCD has  been given. In particular the Hamiltonian of RFT that governs the evolution of physical scattering amplitudes with energy has been derived in two limits - the dilute-dilute limit, where both scattering objects (the projectile and the target) are considered to be small and perturbative (which we refer to as "dilute"), and the dilute-dense limit, where one of the objects is dilute and the other one is "dense", i.e. contains a nonpertubratively large gluonic density. The appropriate evolution in the first limit is given by the BFKL equation\cite{BFKL}, while in the second by the so-called JIMWLK equation\cite{jimwlk,cgc} (and its dual KLWMIJ \cite{klwmij}).
The direct relation between the  JIMWLK and BK  evolution equations \cite{jimwlk,BK}, or Color Glass Condensate (CGC)\cite{cgc} and the RFT  has been recognized in \cite{reggeon}.  
 
 The JIMWLK evolution equation is derived directly from QCD in the leading order perturbative expansion in the dense-dilute regime. As such it does not contain some important effects, like higher order perturbative corrections and the so called Pomeron loops. 
 The NLO corrections to JIMWLK have been derived \cite{nlo} with the conformal part of the kernel known today at the three loop level\cite{caron}.
 
 The hunt after Pomeron loops on the other hand has not concluded yet. The Pomeron loops are important when both, the density effects in the wave function and multiple effects in scattering are equally important.
 Some 15 years ago the activity aimed at incorporating the effects of the  Pomeron loops into the CGC framework has been very lively\cite{pomloops}. Some interesting progress has been made to include both the "splitting" and the "merging" Pomeron  processes into the high energy evolution. This activity unfortunately has not converged to a universally accepted form of high energy evolution and RFT. 
 
 JIMWLK evolution is  valid  only in a limited domain of rapidities, i.e. only as long as one of the colliding objects is dilute.  
 The limitation of the JIMWLK evolution to a dense-dilute scattering is a genuine physical restriction. Even though nominally the JIMWLK equation applies to the evolution of a dense system, the fact that the scattering of this large system is allowed to be perturbative (target is dilute) leads to some paradoxical features. For example, as was anticipated in \cite{KLL} and explicitly demonstrated in \cite{KLLL}, when interpreted as the evolution of QCD wave function of a dense object, JIMWLK evolution leads to appearance of negative probabilities.  The negative probabilities accompany states arising in the evolution with smaller number of gluons than the number of gluons at the outset of the evolution. Physically one expects of course that the number of gluons in the QCD wave function {\bf increases} with energy, while within the JIMWLK framework the number decreases but the low gluon number states appear with negative probability. This strange behavior nevertheless produces correct energy dependence of the S-matrix but only as long as one of the colliding objects is dilute. The violation of unitarity is a precursor of the eventual breakdown of the JIMWLK evolution at high enough energy.
At high energy the Pomeron loops must become important and their effect on the evolution must be significant.

This issue of the unitarity violation in the JIMWLK limit motivates us to reconsider the problem of including Pomeron loops. More precisely  we take up a limited goal to try and extend $H_{JIMWLK}$ in a way that it becomes consistent with 
a very important property of RFT - the self duality. It has been established in \cite{KLduality} that the Hamiltonian that generates the high energy evolution must be invariant under the  dense-dilute duality transformation. Physically the self duality has a very simple meaning. It expresses the fact that a scattering amplitude for a scattering of any two hadrons does not depend on which one of them is right moving and which one is left moving, i.e. which one of them we call the target and which one the projectile.
   As discussed many times in the literature, the JIMWLK evolution explicitly violates the self duality property which one expects to hold in  RFT, since within the domain of validity of JIMWLK the target and the projectile are very different  and thus are explicitly treated differently in  $H_{JIMWLK}$. 
   
   Although self duality alone may not be sufficient to restore unitarity of the evolution, in a zero dimensional toy model addressed in \cite{KLL} it was shown that the unitary Hamiltonian is indeed seld-dual.  Motivated by this, in the present paper we explore possible generalization of the JIMWLK Hamiltonian which restores self duality. Our approach here does not rely on direct derivation from QCD, but instead is akin to typical effective field theory (EFT) attitude: identify relevant degrees of freedom and impose appropriate  symmetries. We also require that in the dense-dilute limit the Hamiltonian reproduces both $H_{JIMWLK}$ and $H_{KLWMIJ}$. We find a family of such Hamiltonians which all reduce to $H_{JIMWLK}$ in the dense-dilute limit and are self dual. We note that one of these Hamiltonians is similar in structure to the so called "diamond action" introduced some years ago in \cite{diamond} and discussed in \cite{Balitsky05}. However a more detailed analysis presented below shows that our construction does not support the condition imposed on the product of Wilson loops in \cite{diamond}, which was crucial in the approach of \cite{diamond} to maintain self duality. Thus our current suggestion is not equivalent to the diamond action of \cite{diamond}. Additionally we note that our  approach relies on the development of RFT formalizm in \cite{KLLL}, and thus provides directly an algorithm for calculation of scattering amplitudes once the Hamiltonian $H_{RFT}$ is specified.

We thus find a family of self-dual RFT Hamiltonians that reproduces all the known limits. Unfortunately it turns out to be technically involved  to check whether the evolution generated by these Hamiltonians is unitary and we are currently unable to answer this question. We are nevertheless encouraged by many similarities with the zero dimensional toy model where the very analogous construction provided a solution to the unitarity problem. The quantitative analysis of this question is left for further research.

The plan of this paper is as follows. In Section 2 we recap the formulation of RFT, its algebra of operators and Hilbert space structure discussed in \cite{KLLL}. In Section 3 we present the construction of $H_{RFT}$  imposing the discrete symmetries of $H_{JIMWLK}$ in addition to self duality. In Section 4 we show that in the dense-dilute limit our $H_{RFT}$ reproduces the JIMWLK and KLWMIJ evolutions. In Section 5 we discuss the continuous symmetries of $H_{RFT}$. This discussion is perturbative, and we conclude that the continuous symmetry group of our $H_{RFT}$ is somewhat surprisingly $SU(N)\times SU(N)\times SU(N)$ \footnote{We have abused the notation here somewhat. The symmetry group is not in fact a direct product of three factors  of $SU(N)$. The more appropriate way to characterize it is to say that the generators contain three linearly independent sets of generators of $SU(N)$. The commutation relations between some of these generators are quite complicated to calculate and thus the full group structure is  not  known. We will expand on this in the body of the paper.}. In Section 6 we consider the relation with the diamond action\cite{diamond}, and show that the so called "diamond condition" on the Wilson lines is violated at second order in $g$. We conclude with discussion in Section 7.


\section{The Reggeon Field Theory: scattering amplitudes and field algebra.}
In this section we briefly recap the general formulation of the Hamiltonian Reggeon Field Theory given in \cite{KLLL}. 

Consider an $S$ matrix element $S_{fi}$ for scattering from the initial QCD state\\ $|\Psi_i\rangle=|\mathbf{x}_1,a_1;...;\mathbf{x}_N,a_N\rangle_T|\mathbf{y}_1, c_1;...;\mathbf{y}_{M}, c_{M}\rangle_P$ to the final state $|\Psi_f\rangle=|\mathbf{x}_1,b_1;...;\mathbf{x}_N,b_N\rangle_T|\mathbf{y}_1, d_1;...;\mathbf{y}_{M}, d_{M}\rangle_P$. Here 
the target state (subscript $T$) contains N gluons, and the projectile state (subscript $P$) contains $M$ gluons. The states are labeled by the transverse coordinates and color indexes of the gluons. At high energy in the eikonal approximation this is given by
\begin{equation}\label{sif}
S_{if}\,\equiv\,\langle \Psi_i| \hat S| \Psi_f\rangle\,=\,\langle L| U^{a_1b_1}(\mathbf{x}_1) \ldots U^{a_Nb_N}(\mathbf{x}_N)\bar{U}^{c_1d_1}(\mathbf{y}_1) \ldots \bar{U}^{c_Md_M}(\mathbf{y}_M)|R\rangle
\end{equation}
where  the left and right RFT Fock vacuum states satisfy 
\begin{equation}\label{fock}
\langle L|\bar U_{ab}=\delta_{ab}\langle L|; \ \ \ \ \ \ U_{ab}|R\rangle=\delta_{ab}|R\rangle.
\end{equation}
The projectile and target  adjoint Wilson line operators are defined in terms of the projectile color charge denstity $\rho^a(\mathbf{x})$ as
\begin{equation}\label{algebra}
\bar U(\mathbf{x})=e^{T^a\frac{\delta}{\delta\rho^a(\mathbf{x})}}\ ; \ \ \ \ \ \ \ \ \ \  U(\mathbf{x})=e^{igT^a\int_y\phi(\mathbf{x-y})\rho^a(\mathbf{y})}
\end{equation}
with
\begin{equation}\label{lorentz}
\alpha^a(\mathbf{x})=\int_{\mathbf{y}}\phi(\mathbf{x-y})\rho^a(\mathbf{y});\ \ \ \ \ \ \ \ \ \ \ \ \phi(\mathbf{x-y})=\frac{g}{2\pi}\ln \frac{|\mathbf{x-y}|}{L}.
\end{equation}
Here $\alpha^a$ is the potential at point $x$ produced by the charge distribution of the target.
The scale $L$ is arbitrary and does not enter calculations of any physical quantities. The $SU(N)$ generators in the adjoint representation are defined in terms of the $SU(N)$ structure constants as
\begin{equation}
T^a_{bc}=-if_{abc}.
\end{equation}

These equations imply non-trivial commutation relations, between $U$ and $\bar U$,  which constitute the algebra of the RFT in analogy with Heisenberg algebra of fields in the ordinary QFT. In order to calculate the scattering amplitude eq.(\ref{sif}) one uses the algebra of $U$ and $\bar U$ to commute  the factors of $U$ to the right of $\bar U$, at which point they disappear by virtue of \eq{fock}.

\begin{figure}[t]   
\centering   
\begin{tabular}{c}
  \includegraphics[width=9cm] {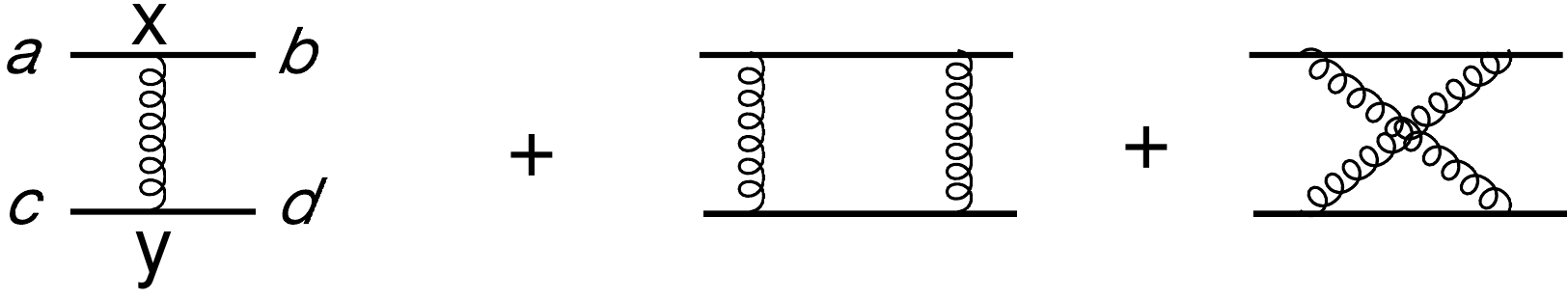} \\ 
  \fig{1}-a\\
  ~~\\
   \includegraphics[width=18cm] {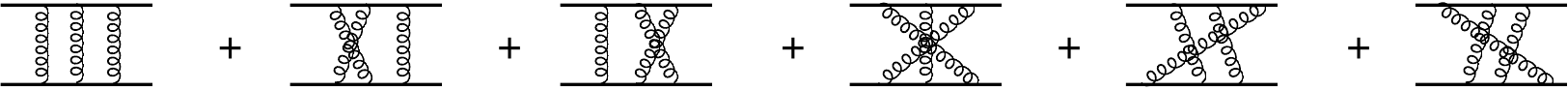}  \\
    \fig{1}-b
         \end{tabular}
\caption{The one, two (\fig{1}-a) and three(\fig{1}-b) gluon exchange contributions to the algebra.}
\label{1}
\end{figure}

This algebra encodes the diagrammatic calculation of scattering amplitudes in the operator language.
Consider for example the scattering of one gluon on one gluon. The scattering amplitude up to second order in $\alpha_s$ is given by 
\begin{equation}
\langle L| U^{ab}(\mathbf{x}) \bar{U}^{cd}(\mathbf{y}) |R\rangle =\delta^{ab}\delta^{cd}- ig\phi(\mathbf{x}-\mathbf{y}) T^i_{ab}T^i_{cd} +\left[ \frac{1}{2!}ig\phi(\mathbf{x}-\mathbf{y})\right]^2  (T^iT^j)_{ab}[(T^iT^j)_{cd} + (T^jT^i)_{cd}]+ \ldots
\end{equation}
This corresponds to the sum of  one and two gluon exchange diagrams in Fig. \ref{1}-a.
In fact as was shown in \cite{KLLL},  higher order terms  organize themselves into all possible diagrams where the relative order of the vertices on the target gluon line is permuted in all possible ways. These are the relevant diagrams for eikonal scattering in the Lorentz gauge. 
The $O(\alpha_s^3)$ contributions correspond to the three gluon exchange diagrams (Fig.\ref{1}-b).

With the algebra encoded in \eq{algebra} and the rule for calculating scattering amplitudes \eq{sif}, the framework of the QCD RFT is defined. To complete the RFT framework  one needs to specify the Hamiltonian  $H_{RFT}$ that generates the evolution of the scattering amplitude in energy. We will spend some time discussing this Hamiltonian below. But before setting along this route  let us recap 
unitarity constraints on any RFT state as derived in \cite{KLLL}. These constraints  must be preserved by energy evolution of the scattering amplitudes. 
This implies a non-trivial  constraint on $H_{\rm RFT}$ \cite{KLLL}.

Eq.(\ref{sif}) is easily extended for scattering of a state which is a superposition of states with fixed number of gluons.
For example, starting with the initial QCD projectile state  
\begin{equation}\label{psii}
|\Psi_i\rangle_P=\sum_{n; \mathbf{x}_i;a_i}C_{a_1,a_2...a_n}|\mathbf{x}_1,a_1;...;\mathbf{x}_n,a_n\rangle
\end{equation}
the eikonal scattering can only produce a state of the form 
\begin{equation}
|\Psi_f\rangle_P=\sum_{n; \mathbf{x}_i;b_i}C_{b_1,b_2...b_n}|\mathbf{x}_1,b_1;...;\mathbf{x}_n,b_n\rangle.
\end{equation}
The same holds for the target
\begin{equation}
|\Psi_i\rangle_T=\sum_{m; \mathbf{y}_j;c_j}\bar C_{c_1,c_2...c_m}|\mathbf{y}_1,c_1;...;\mathbf{y}_m,c_m\rangle
\end{equation}
the eikonal scattering can only produce a state of the form 
\begin{equation}
|\Psi_f\rangle_T=\sum_{m; \mathbf{y}_j;d_i}\bar C_{d_1,d_2...d_m}|\mathbf{y}_1,d_1;...;\mathbf{y}_m,d_m\rangle
\end{equation}
The S-matrix element is given by
\begin{equation}\label{sif1}
S_{if}=\langle L| W_T[U]W_P[\bar U]|R\rangle
\end{equation}
with
\beq\label{wp}
W_P=\sum_{n,\{a,b;\mathbf{x}\}}F^n(\{a,b;\mathbf{x}\})\prod_{i=1}^n[\bar U^{a_ib_i}(\mathbf{x}_i)]
\eeq
\begin{equation}\label{fn}
F^n(\{a,b;\mathbf{x}\})=C_{a_1,a_2...a_n}(\mathbf{x}_1...\mathbf{x}_n)C^*_{b_1,b_2...b_n}(\mathbf{x}_1...\mathbf{x}_n)
\end{equation}
and 
\beq\label{wt}
W_T=\sum_{m,\{c,d;\mathbf{y}\}}\bar F^n(\{c,d;\mathbf{y}\})\prod_{i=1}^m[ U^{c_i,d_i}(\mathbf{y}_i)]
\eeq
\begin{equation}\label{fm}
\bar F^m(\{c,d;\mathbf{y}\})=\bar C_{c_1,c_2...c_m}(\mathbf{y}_1...\mathbf{y}_m)\bar C^*_{d_1,d_2...d_m}(\mathbf{y}_1...\mathbf{y}_m).
\end{equation}
As is obvious from Eqs.(\ref{fn},\ref{fm}), the functions $F$ and $\bar F$ must satisfy the properties of s-channel unitarity \cite{KLLL}
\beq\label{prob}
F^n(\{a,a;\mathbf{x}\})\ge 0; \ \ \ \ \ \  \sum_{n,\{a\}}\int_{\{\mathbf{x}\}}F^n(\{a,a;\mathbf{x}\})=1; \
\eeq
and the same for $\bar F$.

As shown in \cite{KLLL} some these conditions are violated in JIMWLK evolution, which leads to negative probabilities $\bar F$ when evolving the state of a dense target.

\section{The RFT Hamiltonian}

The subject of RFT is the evolution of scattering amplitudes with energy.
In general the energy evolution is generated by the action of the RFT Hamiltonian $H_{RFT}[U,\bar U]$. The S-matrix element of eq.(\ref{sif}) evolved to rapidity $Y$ is given by
\begin{equation}\label{sev}
S_{if}(Y)=\langle L| U^{a_1b_1}(\mathbf{x}_1) \ldots U^{a_Nb_N}(\mathbf{x}_N)e^{YH_{RFT}[U,\bar U]}\bar{U}^{c_1d_1}(\mathbf{y}_1) \ldots \bar{U}^{c_Md_M}(\mathbf{y}_M)|R\rangle.
\end{equation}
\subsection{JIMWLK/KLWMIJ Hamiltonians.}
Exploring the functional form of $H_{RFT}$ is the subject of this paper. Ideally we would like to derive it directly from a QCD calculation. This has been achieved in the  dense-dilute limit, where one of the scattering objects is dense and the other one is dilute. The two versions of the Hamiltonian related by the duality transformation have been derived in \cite{jimwlk,klwmij,cgc}.

When the target is dense and the projectile dilute, the relevant limit is the JIMWLK Hamiltonian:
\begin{equation}\label{hjimwlk}
H_{JIMWLK}=\frac{\alpha_s}{2\pi^2}\int_{\mathbf{x,y,z}}\frac{(\mathbf{x-z})\cdot(\mathbf{y-z})}{ \mathbf{(x-z)^2(y-z)^2}}\left[2\mathcal{J}_L^a(\mathbf{x})\mathcal{J}_R^b(\mathbf{y}) \bar U^{ab}(\mathbf{z})-\mathcal{J}_L^a(\mathbf{x})\mathcal{J}_L^a(\mathbf{y})-\mathcal{J}_R^a(\mathbf{x})\mathcal{J}_R^a(\mathbf{y})\right].
\end{equation}
Here the right and left rotation operators are defined as \cite{ddd}
\begin{equation}\label{j}
\begin{split}
&\mathcal{J}_L^a(\mathbf{x})  =  \left[\frac{1}{2}T^e\frac{\delta}{\delta \rho^e(\mathbf{x})} \left(\coth{\left[\frac{1}{2}T^e\frac{\delta}{\delta \rho^e(\mathbf{x})}\right]}-1\right)\right]^{ba}\rho^b(\mathbf{x})\, ,\\
&\mathcal{J}_R^a(\mathbf{x})  = \left[\frac{1}{2}T^e\frac{\delta}{\delta \rho^e(\mathbf{x})} \left(\coth{\left[\frac{1}{2}T^e\frac{\delta}{\delta \rho^e(\mathbf{x})}\right]}+1\right)\right]^{ba}\rho^b(\mathbf{x})\, ,\\
\end{split}
\end{equation}

The function on the right hand side as usual should be understood as a power series expansion. For a single variable $t$ we have
\begin{equation}
\begin{split}
&M_L(t)\equiv\frac{t}{2}\left(\coth{\frac{t}{2}} -1\right) = \frac{t}{e^t-1} =\sum_{m=0}^{\infty} \frac{B^-_m t^m}{m!} = \sum_{m=0}^{\infty} C^-_m t^m\\
&M_R(t)\equiv\frac{t}{2}\left(\coth{\frac{t}{2}} +1\right) = \frac{t}{1-e^{-t}} =\sum_{m=0}^{\infty} \frac{B^+_m t^m}{m!} =  \sum_{m=0}^{\infty} C^+_m t^m.\\
\end{split}
\end{equation}
Here $B_m^-$ and $B_m^+$ are Bernoulli numbers. They have the properties that $B_{2n}^-=B_{2n}^+$ for all even integers $2n$ while $B_{2n+1}^-=B_{2n+1}^+=0$ for all odd integers $2n+1$ except $B_1^- = -\frac{1}{2} = -B_1^+$. Also the relations $M_L(t) = M_R(t) e^{-t}$ and $M_R(t) = M_L(t) e^{t}$ can be readily verified.

The operators $\mathcal{J}_L^a(\mathbf{x}), \mathcal{J}^a_R(\mathbf{x})$ act as left rotation and right rotation on the Wilson line $\bar{U}^{mn}(\mathbf{x})$, 
\begin{equation}\label{eq:LR_rot_barV}
\begin{split}
&[\mathcal{J}_L^a(\mathbf{x}), \bar{U}^{mn}(\mathbf{y})] = - (T^a \bar{U}(\mathbf{y}))^{mn}\delta(\mathbf{x}-\mathbf{y})\, ,\\
&[\mathcal{J}_R^a(\mathbf{x}), \bar{U}^{mn}(\mathbf{y})] = - ( \bar{U}(\mathbf{y})T^a)^{mn}\delta(\mathbf{x}-\mathbf{y})\,. \\
\end{split}
\end{equation}

One seemingly peculiar feature of these definitions is that when considered as operators on the standard Hilbert space of functions of $\rho$, the operators $\mathcal{J}_{L(R)}$ are not Hermitian
\begin{equation}
\mathcal{J}_L^\dagger\ne \mathcal{J}_L; \ \ \ \ \ \mathcal{J}_R^\dagger\ne \mathcal{J}_R.
\end{equation}
However one has to keep in mind that the operation of Hermitian conjugation of the operators in QCD Hilbert space does not correspond to naive Hermitian conjugation in the RFT space. Without going into detailed discussion here, we refer the reader to \cite{likovner} where it was shown that the RFT transformation that corresponds to Hermitian conjugation  in the QCD Hilbert space is
\begin{equation}
[ {\rm QCD\,\, operator} ]^\dagger\rightarrow(L\leftrightarrow R)^*
\end{equation}
Under this transformation indeed we have 
\begin{equation}
\mathcal{J}_L\rightarrow \mathcal{J}^*_R=\mathcal{J}_L; \ \ \ \ \mathcal{J}_R\rightarrow \mathcal{J}^*_L=\mathcal{J}_R
\end{equation}
as is required for Hermitian operators in the QCD Hilbert space.

The evolution in the reverse situation (dilute target and dense projectile) is governed by the so called KLWMIJ Hamiltonian,
\begin{equation}\label{klwmij}
H_{KLWMIJ}=\frac{\alpha_s}{2\pi^2}\int_{\mathbf{x,y,z}}\frac{(\mathbf{x-z})\cdot(\mathbf{y-z})}{ \mathbf{(x-z)^2(y-z)^2}}\left[2\mathcal{I}_L^a(\mathbf{x})\mathcal{I}_R^b(\mathbf{y}) U^{ab}(\mathbf{z})-\mathcal{I}_L^a(\mathbf{x})\mathcal{I}_L^a(\mathbf{y})-\mathcal{I}_R^a(\mathbf{x})\mathcal{I}_R^a(\mathbf{y})\right]
\end{equation}
where $\mathcal{I}_{L(R)}$ are defined as 
\begin{equation}
\begin{split}
&\mathcal{I}_L^a(\mathbf{x})  = \frac{-i}{g}\frac{\delta}{\delta \alpha^b(\mathbf{x})}  \left[\frac{1}{2}T^eig\alpha^e(\mathbf{x})\left(\coth{\left[\frac{1}{2}T^eig\alpha^e(\mathbf{x})\right]}-1\right)\right]^{ba}\, ,\\
&\mathcal{I}_R^a(\mathbf{x})  = \frac{-i}{g}\frac{\delta}{\delta \alpha^b(\mathbf{x})}  \left[\frac{1}{2}T^eig\alpha^e(\mathbf{x})\left(\coth{\left[\frac{1}{2}T^eig\alpha^e(\mathbf{x})\right]}+1\right)\right]^{ba}\, ,\\
\end{split}
\end{equation}
with $
\alpha^a(\mathbf{x})$ defined in eq.(\ref{lorentz}).  These satisfy
\begin{equation}
\begin{split}
&[U^{mn}(\mathbf{y}), \mathcal{I}^a_L(\mathbf{x})] = - (T^a U(\mathbf{y}))^{mn}\delta(\mathbf{x}-\mathbf{y})\, ,\\
&[U^{mn}(\mathbf{y}), \mathcal{I}^a_R(\mathbf{x})] = - ( U(\mathbf{y})T^a)^{mn}\delta(\mathbf{x}-\mathbf{y})\, .\\
\end{split}
\end{equation}

The two sets of operators satisfy two copies of  $SU(N)\times SU(N)$ commutation relations:
\begin{equation}
\begin{split}
&[\mathcal{J}_L^a(\mathbf{x}), \mathcal{J}_L^b(\mathbf{y})] = if^{abc} \mathcal{J}^c_L(\mathbf{x})\delta(\mathbf{x}-\mathbf{y})\,, \\
&[\mathcal{J}_R^a(\mathbf{x}), \mathcal{J}_R^b(\mathbf{y})] = -if^{abc} \mathcal{J}^c_R(\mathbf{x})\delta(\mathbf{x}-\mathbf{y})\, \\
&[\mathcal{J}_L^a(\mathbf{x}), \mathcal{J}_R^a(\mathbf{y})] =0\, .
\end{split}
\end{equation}

and 
\begin{equation}
\begin{split}
&[\mathcal{I}_L^a(\mathbf{x}), \mathcal{I}_L^b(\mathbf{y})] = -if^{abc} \mathcal{I}^c_L(\mathbf{x})\delta(\mathbf{x}-\mathbf{y})\,, \\
&[\mathcal{I}_R^a(\mathbf{x}), \mathcal{I}_R^b(\mathbf{y})] = if^{abc} \mathcal{I}^c_R(\mathbf{x})\delta(\mathbf{x}-\mathbf{y})\, \\
&[\mathcal{I}_L^a(\mathbf{x}), \mathcal{I}_R^a(\mathbf{y})] =0\, .
\end{split}
\end{equation}
The commutation relations between $\mathcal{J}$ and $\mathcal {I}$ are rather complicated and we will not attempt to derive them here.

The Hamiltonian of RFT must possess a property of self duality, i.e. it has to be invariant under the transformation that interchanges the projectile and the target. This is obvious from the point of view of QCD, since it is immaterial which one of the colliding objects we call the target, and which one the projectile. Thus scattering of an N gluon projectile on an M gluon target is the same as scattering of an M gluon projectile on an N gluon target. The JIMWLK (and likewise KLWMIJ) Hamiltonian is not self dual, since it is only meant to be valid in the very asymmetric regime where one of the colliding objects is dense and one is dilute. This lack of self duality means among other things, that JIMWLK cannot be used at asymptotically high energies, where the projectile becomes dense as well. It is thus clearly desirable to find a self dual extension of $H_{JIMWLK}$.

Some years ago a considerable effort has been dedicated  to a search for a self dual extension of the Hamiltonian. One such extension in the context of large $N_c$ Pomeron theory was suggested by Braun \cite{braun}. The solutions to the Braun theory however exhibit a nonphysical bifurcating behavior \cite{motyka} which  was an original motivation for the study of \cite{KLL}. It was shown in \cite{KLL} that  Braun's theory suffers from unitarity violation. Other attempts based on the QCD path integral approach were reported in \cite{diamond,Balitsky05}. Those works have proposed the so called "diamond action" as a self dual effective action of RFT. Although the question has not been settled, in recent years this effort has only been simmering on a back burner.

Here we return to this problem motivated by considerations of unitarity. As we showed in \cite{KLLL}, the JIMWLK Hamiltonian violates QCD unitarity constraints when acting on the dense target wave function.
In view of the discussion in \cite{KLL} of the zero dimensional toy model, it seems likely that the self duality of $H_{RFT}$  is necessary in order to restore unitarity.
In this section we present a self dual $H_{RFT}$ and show that it reduces to $H_{JIMWLK}$ and $H_{KLWMIJ}$ in the appropriate dense-dilute limit.

\subsection{The self dual extension}
\subsubsection{The symmetries}
Our strategy in this paper is similar to that of  EFT: we are not going to attempt to derive $H_{RFT}$ from first principles, but will rather construct a family of Hamiltonians which on the one hand reduce to $H_{JIMWLK}$ and $H_{KLWMIJ}$ in the appropriate limits, and on the other hand are symmetric under the known symmetries of $H_{JIMWLK}$ in addition to being self dual. 

The symmetries of $H_{JIMWLK}$ have been analyzed for example in \cite{reggeon} and \cite{yin}.
$H_{JIMWLK}$ possesses the continuous symmetry group $SU_L(N)\times SU_R(N)$ generated by $\mathcal{J}_{L(R)}$. In addition it has the discrete $Z_2^S\times Z_2^C$ symmetry group with the two discrete transformations acting in the following way:

1. The signature $Z_2^S$
\begin{equation}\label{signature}
SUS^\dagger=U^\dagger; \ \ \ \ S\bar US^\dagger=\bar U^\dagger;\ \ \ \ \ \ S\mathcal{J}_LS^\dagger=-\mathcal{J}_R; \ \ \ \ S\mathcal{I}_LS^\dagger=-\mathcal{I}_R.
\end{equation}

2. The charge conjugation $Z_2^C$.

For simplicity we choose to work in the basis where the generators in the fundamental representation $t^a$ are either real and symmetric or imaginary and antisymmetric. In this basis the charge conjugation symmetry corresponds to changing the sign of the real generators since this has the effect $t^a\rightarrow -t^{a*}$ which interchanges the generators in fundamental and anti fundamental representations.
Defining the matrix
\begin{equation}
c_{ab}=-2tr[t^at^{*b}]
\end{equation}
the "second quantized" form of the transformation is
\begin{equation}\label{cc}
C\mathcal{J}^a_{L(R)}C^\dagger=c^{ab}\mathcal{J}^b_{L(R)};\ \ \ \ \ C\mathcal{I}^a_{L(R)}C^\dagger=c^{ab}\mathcal{I}^b_{L(R)}.
\end{equation}
The eikonal factors in fundamental ($U_F$) and adjoint ($U$) representations transform as
\begin{equation}
 CU_FC^\dagger=U_F^*;\ \ \ \ C\bar U_FC^\dagger=\bar U_F^* .
 \end{equation}
\begin{equation}
CU_{ab}C^\dagger=c_{ac}U_{cd}c_{db}; \ \ \ \ C\bar U_{ab}C^\dagger=c_{ac}\bar U_{cd}c_{db}.\ \ \ \ \end{equation}

We expect both the discrete symmetries of $H_{JIMWLK}$ to remain the symmetries of the general $H_{RFT}$ since they directly reflect the symmetries of QCD. The situation with $SU_L(N)\times SU_R(N)$ is less clear. It is certainly true that we expect the diagonal vector subgroup $SU_V(N)$ to be a symmetry of $H_{RFT}$, since it descends directly from the global color group of QCD as it rotates simultaneously the initial and final scattering states. The left rotation acts only on the initial states and may be an accidental symmetry of the dense-dilute limit. Thus we will not insist on $SU_L(N)$ and $SU_R(N)$ to be separate symmetries but will return to this question later.

In addition to these symmetries which are symmetries of $JIMWLK$ limit, we will also require $H_{RFT}$ to be invariant under the  dense dilute duality $Z_2^D$. To understand how the duality transformation  acts on the field variables in the current RFT setup, we recall that physically it simply interchanges the projectile and the target. In other words for basic scattering amplitude we should have
\begin{eqnarray}
&&\langle L| U^{a_1b_1}(\mathbf{x}_1) \ldots U^{a_Nb_N}(\mathbf{x}_N)\bar{U}^{c_1d_1}(\mathbf{y}_1) \ldots \bar{U}^{c_Md_M}(\mathbf{y}_M)|R\rangle\,\,\rightarrow\\
&&~~~~~~~~~~~~~~~~~~~~~~~~~~~~~~~~~~~~~~~~~~~~ \langle L| {U}^{d_1c_1}(\mathbf{y}_1) \ldots {U}^{d_Mc_M}(\mathbf{y}_M)\bar U^{b_1a_1}(\mathbf{x}_1) \ldots \bar U^{b_Na_N}(\mathbf{x}_N)|R\rangle.\nonumber
\end{eqnarray}
Self duality, or invariance, under $Z^D_2$ is a realization of the fact that the two amplitudes must be equal at any collision enery
\begin{eqnarray}
&&\langle L| U^{a_1b_1}(\mathbf{x}_1) \ldots U^{a_Nb_N}(\mathbf{x}_N)\bar{U}^{c_1d_1}(\mathbf{y}_1) \ldots \bar{U}^{c_Md_M}(\mathbf{y}_M)|R\rangle\\
&&~~~~~~~~~~~~~~~~~~~~~~~~~~~~~~~~~~~~~~~~~=~~ \langle L| {U}^{d_1c_1}(\mathbf{y}_1) \ldots {U}^{d_Mc_M}(\mathbf{y}_M)\bar U^{b_1a_1}(\mathbf{x}_1) \ldots \bar U^{b_Na_N}(\mathbf{x}_N)|R\rangle.\nonumber
\end{eqnarray}
When considered as a transformation acting on a function of the basic fields $\rho$ and $\frac{\delta}{\delta\rho}$, the $Z^D_2$  transformation can be written as
\begin{equation}\label{dualityt}
F[\rho,\frac{\delta}{\delta\rho}]\rightarrow F^\dagger[-\frac{i}{g} \frac{\delta}{\delta \alpha^a},ig\alpha^a].
\end{equation}
In terms of individual operators this is \cite{KLduality}
\begin{eqnarray}
&&\rho^a \rightarrow -\frac{i}{g} \frac{\delta}{\delta \alpha^a}\, ,\qquad  \frac{\delta}{\delta \rho^a}\rightarrow-ig\alpha^a\\
&&U\rightarrow \bar U; \ \ \ \ \ \ \ \ \mathcal{J}_{L(R)}\rightarrow \mathcal{I}^\dagger_{R(L)}.\nonumber
\end{eqnarray}
 However, in addition to this action one has to take an overall Hermitian conjugation of the whole expression which is being transformed. Note that due to this additional action of Hermitian conjugation the duality transformation  $Z^D_2$        
   cannot be represented by an action of a unitary operator on the RFT Hilbert space. This is similar to time reversal in quantum mechanics, which is not a unitary but an anti unitary transformation. Recall that anti unitary transformation involves complex conjugation of an operator function in addition to the transformation of basic variables. The duality is not an anti unitary transformation either, since it involves hermitian conjugation rather than a simple complex conjugation of a function $F$. Nevertheless, just like the time reversal in quantum mechanics, it is a bona fide linear transformation in the Hilbert space and thus should be considered on par with other symmetries of the theory.
   
\subsubsection{The "left" and "right" Wilson lines}

To construct $H_{RFT}$ 
let us introduce the following Wilson line like operators  in the \textit{fundamental representation}
\begin{equation}\label{fundv}
\begin{split}
&V_L(\mathbf{x}) = \mathrm{Exp}\left\{i\int_{\mathbf{y}} g\phi(\mathbf{x}-\mathbf{y}) t^e \mathcal{J}^e_L(\mathbf{y})\right\}\\
&V_R(\mathbf{x}) =\mathrm{Exp}\left\{-i\int_{\mathbf{y}}g \phi(\mathbf{x}-\mathbf{y}) t^e \mathcal{J}^e_R(\mathbf{y})\right\}\\
&\bar{V}_L(\mathbf{x}) = \mathrm{Exp}\left\{i\int_{\mathbf{y}} g\phi(\mathbf{x}-\mathbf{y}) t^e \mathcal{I}^e_L(\mathbf{y})\right\}\\
&\bar{V}_R(\mathbf{x}) = \mathrm{Exp}\left\{-i\int_{\mathbf{y}} g\phi(\mathbf{x}-\mathbf{y}) t^e \mathcal{I}^e_R(\mathbf{y})\right\}.\\
\end{split}
\end{equation}
These expressions resemble our reggeized gluon operators $U$ and $\bar U$. However they are defined in terms of $SU(N)$ generators $\mathcal{J}_{L(R)}$ and $\mathcal{I}_{L(R)}$ rather than commuting variables $\rho$. 

The reason to introduce these operators is that they look like appropriate building blocks for $H_{RFT}$. Recall that we need $H_{RFT}$ to reduce to $H_{JIMWLK}$     in the dense dilute limit, i.e. in the leading order of expansion in powers of $\rho$. Now $H_{JIMWLK}$ is a simple function when written in terms of $\mathcal{J}_{L(R)}$ rather than the regular Wilson line operators $U$. It therefore seems likely that in order to extend it beyond the dense-dilute limit the basic building blocks also should be simple function of $\mathcal{J}$'s. On the other hand $H_{JIMWLK}$ is  also a simple function of $\bar U$. Given that we want to impose self duality on $H_{RFT}$ it is reasonable to choose our building blocks to be in some way similar to Wilson lines. Hence the motivation to introduce the operators in \eq{fundv}. We chose to discuss these operators in fundamental representation for simplicity. As we will show later, the construction we propose works with an arbitrary representation of $SU(N)$, thus providing an infinite set of Hamiltonians that satisfy our requirements.

\begin{figure}[tbp]
\centering 
\includegraphics[width=.5\textwidth]{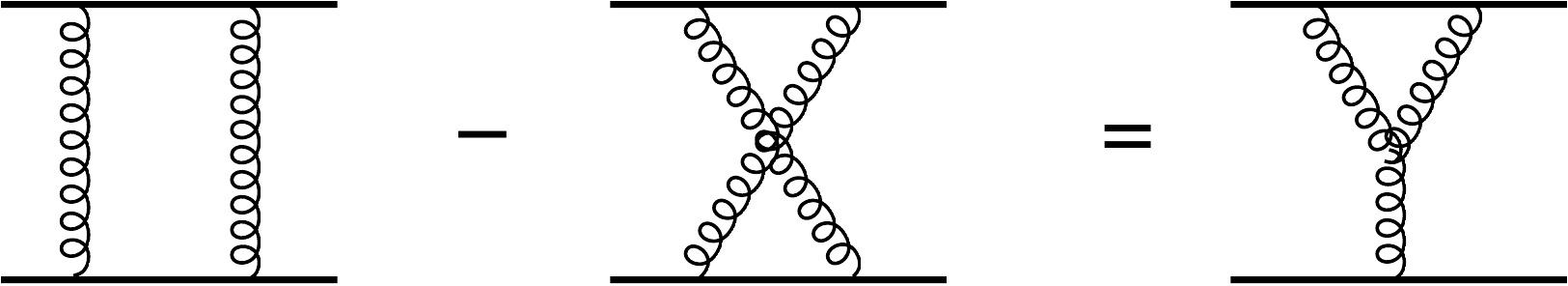}
\caption{Reggeization    corrections to a single gluon exchange.}
\label{2}
\end{figure}

When calculating the RFT ``correlators" of these operators with $U$ and $\bar U$, the ordering of the vertices is important, unlike in the calculation of correlators of $U$'s and $\bar U$'s among themselves. For example consider the simplest correlator
\begin{equation}
\langle L|V^{\alpha\beta}_L(\mathbf{x})\bar U^{cd}(\mathbf{y})|R\rangle=\delta^{\alpha\beta}\delta^{cd}-ig\phi(\mathbf{x}-\mathbf{y}) t^i_{\alpha\beta}T^i_{cd} +\frac{1}{2!}\left[ ig\phi(\mathbf{x}-\mathbf{y})\right]^2  (t^it^j)_{\alpha\beta}(T^jT^i)_{cd} + \ldots
\end{equation}
where the ellipsis denotes contributions of order $g^6$ and higher, i.e. three and higher gluon exchange diagrams.
For comparison, a similar correlator for the fundamental Wilson line  defined as   
$$V(\mathbf{x}) =  U_F(\mathbf{x})\,=\,\mathrm{Exp}\left\{i\int_{\mathbf{y}} \phi(\mathbf{x}-\mathbf{y}) t^e \rho^e(\mathbf{y})\right\}$$
 is
\begin{equation}
\langle L|V^{\alpha\beta}(\mathbf{x})\bar U^{cd}(\mathbf{y})|R\rangle=\delta^{\alpha\beta}\delta^{cd}-ig\phi(\mathbf{x}-\mathbf{y}) t^i_{\alpha\beta}T^i_{cd} +\left[ \frac{1}{2!}ig\phi(\mathbf{x}-\mathbf{y})\right]^2  (t^it^j)_{\alpha\beta}[(T^iT^j)_{cd} +(T^jT^i)_{cd}] + \ldots
\end{equation}
At the two gluon exchange level the difference between the two is
\begin{equation}
\langle L|V^{\alpha\beta}_L(\mathbf{x})\bar U^{cd}(\mathbf{y})|R\rangle-\langle L|V^{\alpha\beta}(\mathbf{x})\bar U^{cd}(\mathbf{y})|R\rangle=-\frac{1}{4}\left[ ig\phi(\mathbf{x}-\mathbf{y})\right]^2  (t^it^j)_{\alpha\beta}[T^i,T^j]_{cd} 
\end{equation}
which corresponds to the diagram in \fig{2}. Note that this difference is a two gluon exchange in the octet channel, and may be viewed simply as the reggeization correction to a single gluon exchange. 

\begin{figure}[tbp]
\centering 
\includegraphics[width=.5\textwidth]{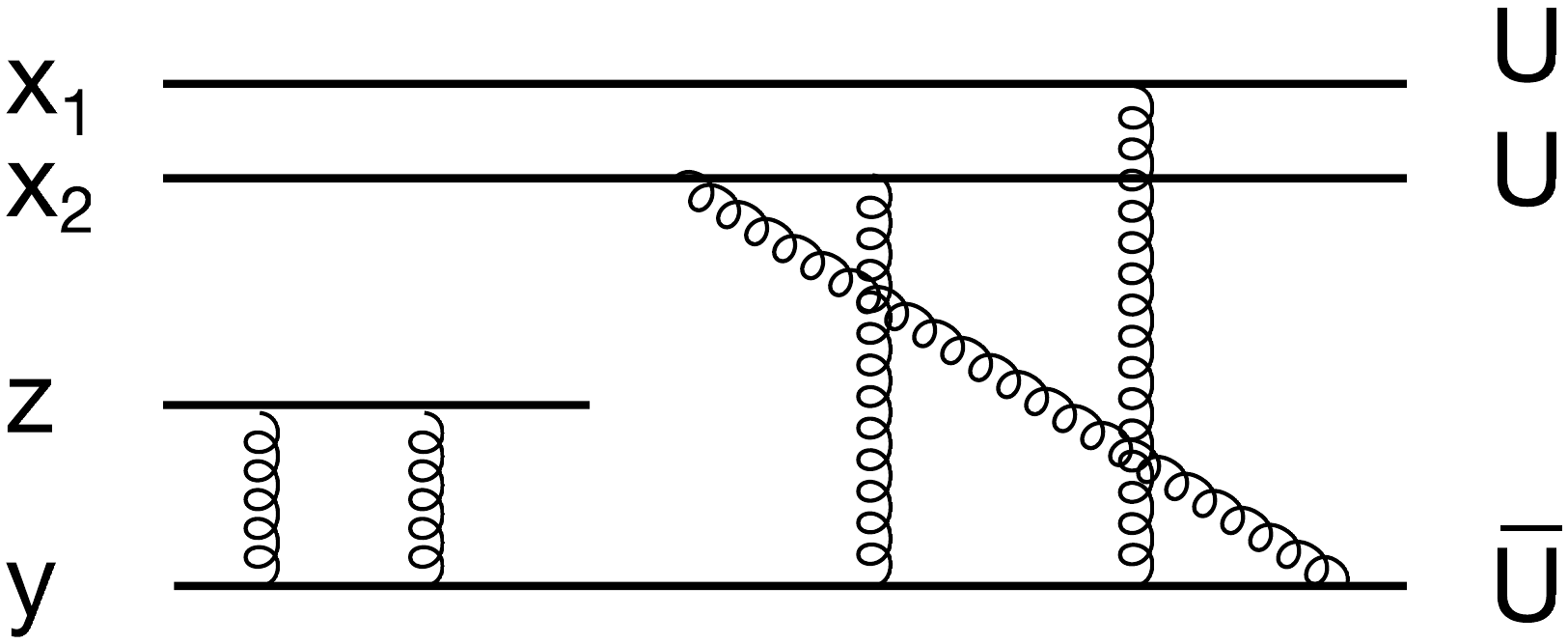}
\caption{A sample diagram for a correlator that includes $V_L$. }
\label{3}
\end{figure}

In general if one thinks about $V_L$ as representing a fundamentally charged parton in the target wave function, the parton in question would be something of a black sheep. It would always scatter on the projectile only after all the other partons have had their day. As an example, a sample diagram corresponding to the calculation of the correlator $\langle L|U(\mathbf{x_1})U(\mathbf{x_2})V_L(\mathbf{z})\bar U(\mathbf{y})|R\rangle$ is depicted on \fig{3}. Note that all the gluons  exchanged between $\bar U$ and $V_L$ attach to the $\bar U$ line to the left of any gluon exchanged between $\bar U$ and any of the $U$'s. This follows since  $V_L$ contains only left rotation generators of $\bar U$.
Similarly, $V_R$ only contains right rotation operators, and therefore in a scattering diagram always exchanges gluons with the projectile {\bf before} any other exchanges with target gluons.

Also note that operatorially $V_L$ and $V_R$ {\bf do not commute} with $U$, although they commute with each other. 
Similar comments apply to $\bar V_{L(R)}$.


\subsubsection{Constructing $H_{RFT}$}
\begin{figure}[tbp]
\centering 
\includegraphics[width=.35\textwidth]{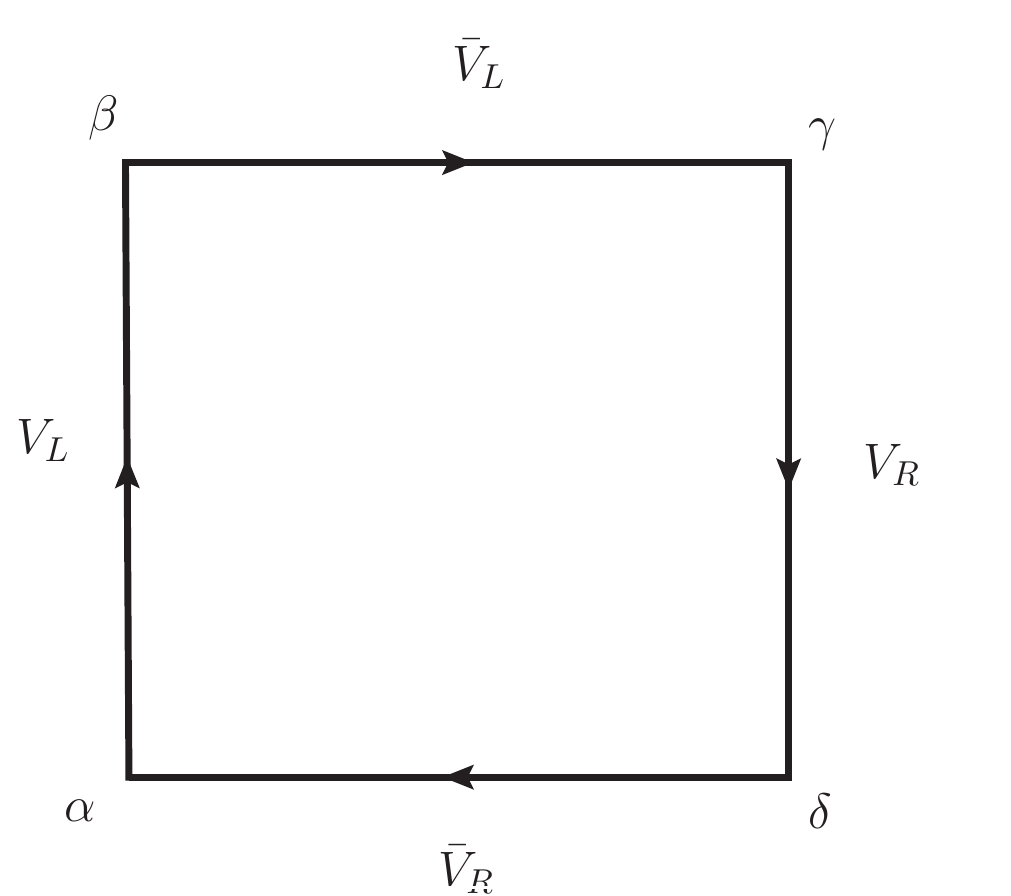}
\caption{The Reggeon field theory Hamiltonian $H^{(1)}_{RFT}$. The arrows indicate the directions of color charge flow. }
\label{4}
\end{figure}

Let us now consider the following expression
\begin{equation}\label{eq:RFT_Hamiltonian_v1}
\begin{split}
H^{(1)}_{RFT} =&\frac{1}{\pi g^2} \int d^2\mathbf{x} \mathrm{Tr}[\partial^2V_L(\mathbf{x})\bar{V}_L(\mathbf{x})V_R(\mathbf{x})\bar{V}_R(\mathbf{x}) +  V_L(\mathbf{x})\bar{V}_L(\mathbf{x})\partial^2 V_R(\mathbf{x})\bar{V}_R(\mathbf{x})\\
&\qquad\qquad \qquad +2 \partial_iV_L(\mathbf{x})\bar{V}_L(\mathbf{x})\partial_i V_R(\mathbf{x})\bar{V}_R(\mathbf{x})]\\
=&\frac{1}{\pi g^2}\int d^2\mathbf{x} \, \bar{V}_L^{\beta\gamma}(\mathbf{x}) \bar{ V}^{\delta\alpha}_R(\mathbf{x}) \partial^2 [V_L^{\alpha\beta}(\mathbf{x}) V_{R}^{ \gamma\delta}(\mathbf{x})]\, \\
=&\frac{1}{\pi g^2}\int d^2\mathbf{x} \, \partial^2[\bar{V}_L^{\beta\gamma}(\mathbf{x}) \bar{ V}^{ \delta\alpha}_R(\mathbf{x})] V_L^{\alpha\beta}(\mathbf{x}) V_{R}^{ \gamma\delta}(\mathbf{x})\, 
\end{split}
\end{equation}
where in the last line we have integrated by parts assuming that the boundary terms vanish.
Note that the order of factors is important, since the operators $V$ and $\bar V$ do not commute with each other.
In \eqref{eq:RFT_Hamiltonian_v1}  all factors $V_L, V_R$ are understood as positioned to the right of any factor $\bar{V}_L, \bar{V}_R$. The diagram that schematically represents the color flow between the four Wilson lines is shown in \fig{4}. 

We start with this  expression since  as we will see shortly it reproduces both, the JIMWLK and the KLWMIJ Hamiltonians in the appropriate dense-dilute limit. 
Following our EFT like strategy we would like to impose on $H_{RFT}$ the discrete symmetries discussed above. It turns out that it is quite easy to do. 

We start with the duality transformation $Z_2^D$. We perform the transformation in two steps. First we perform the canonical transformation
\begin{equation}
\rho^a \leftrightarrow -\frac{i}{g} \frac{\delta}{\delta \alpha^a}\, ,\qquad \frac{\delta}{\delta \rho^a} \leftrightarrow -ig \alpha^a
\end{equation}
under which 
\begin{equation}
\mathcal{J}_L^a \leftrightarrow \mathcal{I}_R^{a\dagger}, \qquad \mathcal{J}_R^a \leftrightarrow \mathcal{I}_L^{a\dagger}
\end{equation}
or equivalently, 
\begin{equation}\label{eq:self_duality_one}
\begin{split}
&V_L\rightarrow \bar{V}^{\dagger}_R, \qquad V_R\rightarrow \bar{V}^{\dagger}_L.\\
&\bar{V}_L \rightarrow V_R^{\dagger}, \qquad \bar{V}_R \rightarrow  V_L^{\dagger}.\\
\end{split}
\end{equation}
Second, in accordance with \eq{dualityt}  we take the Hermitian conjugation of the transformed Hamiltonian to obtain
\begin{equation}
\begin{split}
H_{RFT}^{(1) dual } = & \left(\frac{1}{\pi g^2} \int d\mathbf{x}\, V_R^{\dagger \beta\gamma}(\mathbf{x}) V^{\dagger \delta\alpha}_L(\mathbf{x}) \partial^2 [\bar{V}_R^{\dagger \alpha\beta}(\mathbf{x})\bar{V}_L^{\dagger \gamma\delta}(\mathbf{x})]\right)^{\dagger} \\
=&\frac{1}{\pi g^2} \int d\mathbf{x}\,  \partial^2 [\bar{V}_R^{\beta\alpha}(\mathbf{x})\bar{V}_L^{\delta\gamma}(\mathbf{x})]V_R^{\gamma\beta}(\mathbf{x}) V^{\alpha\delta}_L(\mathbf{x})=H_{RFT}^{(1)}.\\
\end{split}
\end{equation}
Thus we find that $H_{RFT}^{(1)}$ is self dual already.

The next in line is the signature transformation \eq{signature}
\begin{equation}
\begin{split}
& V_L \leftrightarrow V_R,\qquad  \bar{V}_L \leftrightarrow \bar{V}_R,\\
\end{split}
\end{equation}
It is easily seen that $H^{(1)}_{RFT}$ is invariant under this transformation.

The only remaining discrete symmetry is charge conjugation. Although $H^{(1)}_{RFT}$ itself is not 
invariant it is easy to rectify this.

According to \eq{cc} the charge conjugation transformation acts on the left and right Wilson lines.
From the definition of $\mathcal{J}_{L(R)}^a$ and $\mathcal{I}_{L(R)}^a$, taking complex conjugate, one obtains
\begin{equation}
\begin{split}
&CV_LC^\dagger\equiv V_L^{c} = \mathrm{exp}\left\{-ig\int_{\mathbf{y}} \phi(\mathbf{x}-\mathbf{y})t^{e\ast} \mathcal{J}_L^e(\mathbf{y})\right\},\\
&CV_RC^\dagger\equiv V_R^{c} = \mathrm{exp}\left\{ig\int_{\mathbf{y}} \phi(\mathbf{x}-\mathbf{y})t^{e\ast} \mathcal{J}_R^e(\mathbf{y})\right\},\\
&C\bar V_LC^\dagger\equiv \bar{V}_L^{c} = \mathrm{exp}\left\{-ig\int_{\mathbf{y}} \phi(\mathbf{x}-\mathbf{y})t^{e\ast} \mathcal{I}_L^e(\mathbf{y})\right\},\\
&C\bar V_RC^\dagger\equiv \bar{V}_R^{c} = \mathrm{exp}\left\{ig\int_{\mathbf{y}} \phi(\mathbf{x}-\mathbf{y})t^{e\ast} \mathcal{I}_R^e(\mathbf{y})\right\}.\\
\end{split}
\end{equation}
Applying the charge conjugation on $H^{(1)}_{RFT}$ we obtain
\begin{equation}
CH^{(1)}_{RFT}C^\dagger\equiv H^{(1)c}_{RFT} = \frac{1}{\pi g^2} \int d\mathbf{x} \, \bar{V}_L^{c, \beta\gamma}(\mathbf{x}) \bar{V}^{c,\delta\alpha}_R(\mathbf{x})\partial^2\left[ V_L^{c, \alpha\beta}(\mathbf{x})V_{R}^{c,\gamma\delta}(\mathbf{x})\right]. \
\end{equation}
It is easy to see that $H^{(1)c}_{RFT}$ is by itself invariant under the signature and duality transformations.
Therefore, the following Hamiltonian is invariant under all relevant discrete symmetries:
\begin{equation}\label{eq:correct_RFT}
\begin{split}
H_{RFT} = &\frac{1}{2} \left(H^{(1)}_{RFT} + H^{(1)c}_{RFT}\right)\\
=& \frac{1}{2\pi g^2} \int d^2\mathbf{x} \left( \bar{V}_L^{\beta\gamma}(\mathbf{x}) \bar{V}^{ \delta\alpha}_R(\mathbf{x}) \partial^2 \left[V_L^{\alpha\beta}(\mathbf{x})V_{R}^{\gamma\delta}(\mathbf{x})\right] +\bar{V}_L^{c, \beta\gamma}(\mathbf{x}) \bar{V}^{c,\delta\alpha}_R(\mathbf{x})\partial^2\left[ V_L^{c, \alpha\beta}(\mathbf{x})V_{R}^{c,\gamma\delta}(\mathbf{x})\right] \right). \end{split}
\end{equation}

So far we have not discussed the continuous symmetries of $H_{RFT}$. We will postpone this discussion to Section 5 after we consider the dense-dilute limit.

We have found a candidate RFT Hamiltonian which is self dual. In fact the construction above defines a family of self dual Hamiltonians. In particular rather than using the fundamental representation for defining $V_{L(R)}$ and $\bar V_{L(R)}$ we could have used any representation of the color group. Any one of these variations is self dual and, as we will see later reduces to the JIMWLK Hamiltonian in the dense-dilute limit. We do not have any a priori reason to prefer one of these versions to another, although it may seem unnatural to involve very high representations of the color group. One should also note that for representations that have vanishing $N$-ality, like the adjoint representation one has  $H^{(1)}_{RFT} =H^{(1)c}_{RFT}$ which is a  simplifying feature. 

In this paper we will  be working with the fundamental representation defined in \eq{fundv} when deriving the JIMWLK and KLWMIJ limits so that not to loose generality. We will show that $H^{(1)}_{RFT}$ and $  H^{(1)c}_{RFT}$ separately reduce to $H_{JIMWLK}$ and $H_{KLWMIJ}$ in appropriate limits and that this feature extends to any representation of $SU(N)$.

\section{The dense-dilute limit.}
The most important test for $H_{RFT}$ is that it must reproduce $H_{JIMWLK}$ in the dense-dilute limit. In this section we demonstrate explicitly that this is indeed the case. 

The dense-dilute limit arises when the number of gluons in the projectile is of order one, while the number of gluons in the target is large, parametrically $n\sim O(1/\alpha_s^2)$. Thus we are considering the amplitude in \eq{sif} and \eq{sev} where the number of factors $\bar U$ is of order one, and the number of factors $U$ is of order $1/\alpha_s^2$.
In this limit several simplifications occur.

We will first give a simplified argument, and then complete the mathematical details of the demonstration.

\begin{figure}[tbp]
\centering 
\includegraphics[width=.4\textwidth]{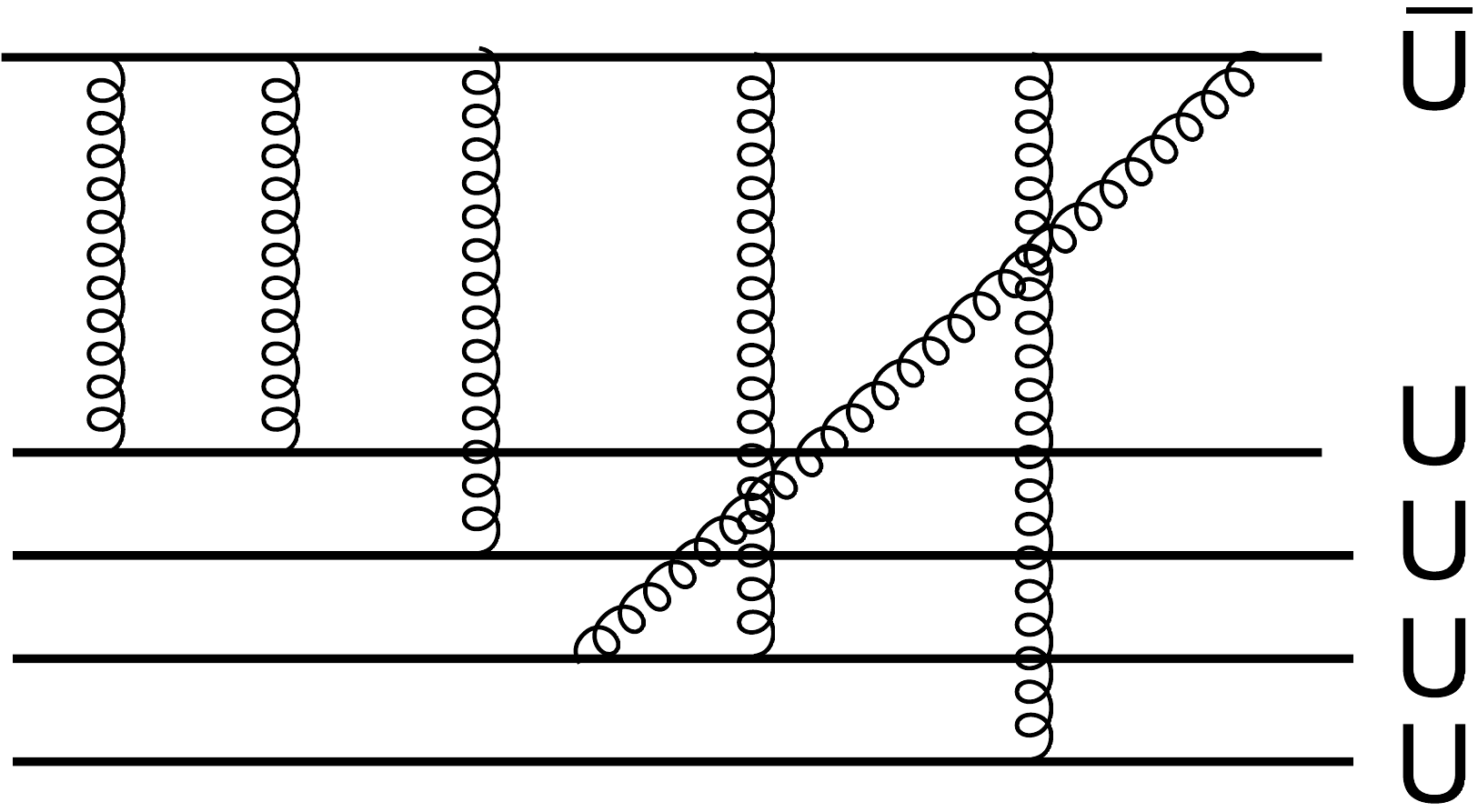}
\caption{A representative diagram for scattering of a single projectile gluon in the dense-dilute limit. Here we have drawn the projectile gluon on the top of the figure and the target gluons at the bottom.}
\label{5}
\end{figure}

First of all, note that at weak coupling any given projectile gluon can exchange at most two gluons with any given target gluon. However,  since the number of gluons in the target is large, a projectile gluon can multiply scatter on many gluons of the target. 
A representative diagram for scattering of a single projectile gluon is depicted on \fig{5}.
The diagram in \fig{5} contains single and double gluon exchanges between individual pairs of gluons.
If a single gluon exchange is present such a diagram contributes to an inelastic amplitude as the final state of the scattering process is necessarily different from the initial state. The elastic amplitude has contribution only from two gluon exchanges where the two gluons are in the color singlet. Since every two gluon exchange carries a factor $\alpha_s^2$, and there are in total $O(1/\alpha_s^2)$ target partons that can participate in the scattering, the total elastic scattering amplitude in the dense-dilute limit  is of order unity\footnote{
Single gluon exchanges behave a little differently. One does not add single gluon exchange amplitudes between a given projectile gluon and different target gluons since those lead to different final states of the target and do not contribute to the same $S$ matrix element. Instead the single gluon exchanges with distinct target gluons lead to appearance of many nonvanishing off diagonal matrix elements of the $S$ matrix albeit each such matrix element is of order $\alpha_s$.  The number of such nonvanishing matrix elements is $O(1/\alpha_s^2)$.}.

\begin{figure}[tbp]
\centering 
\includegraphics[width=.7\textwidth]{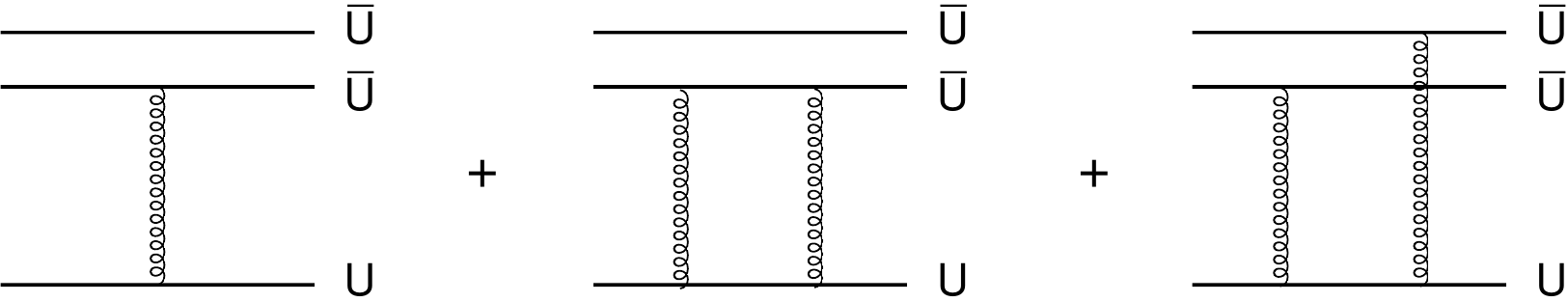}
\caption{A representative diagram for scattering of a single target gluon in the dense-dilute limit. Here we have drawn the target gluon on the bottom of the figure and the projectile gluons at the top.}
\label{6}
\end{figure}

On the other hand since the projectile is dilute, every target gluon can only scatter either on one or two projectile gluons. The appropriate diagrams are represented on \fig{6}. 
Technically this means that in the dense-dilute limit all factors of $U$ have to be expanded to second order in 
$\rho$. This insures that once two gluons are exchanged between a target gluon and the projectile, the target gluon does not participate in any further scattering.

Now consider the diagrams as in \fig{6} but which, instead of one of the factors $\bar U$ contain a factor $\bar V_L$ that appears in the RFT Hamiltonian. 

\begin{figure}[tbp]
\centering 
\includegraphics[width=.3\textwidth]{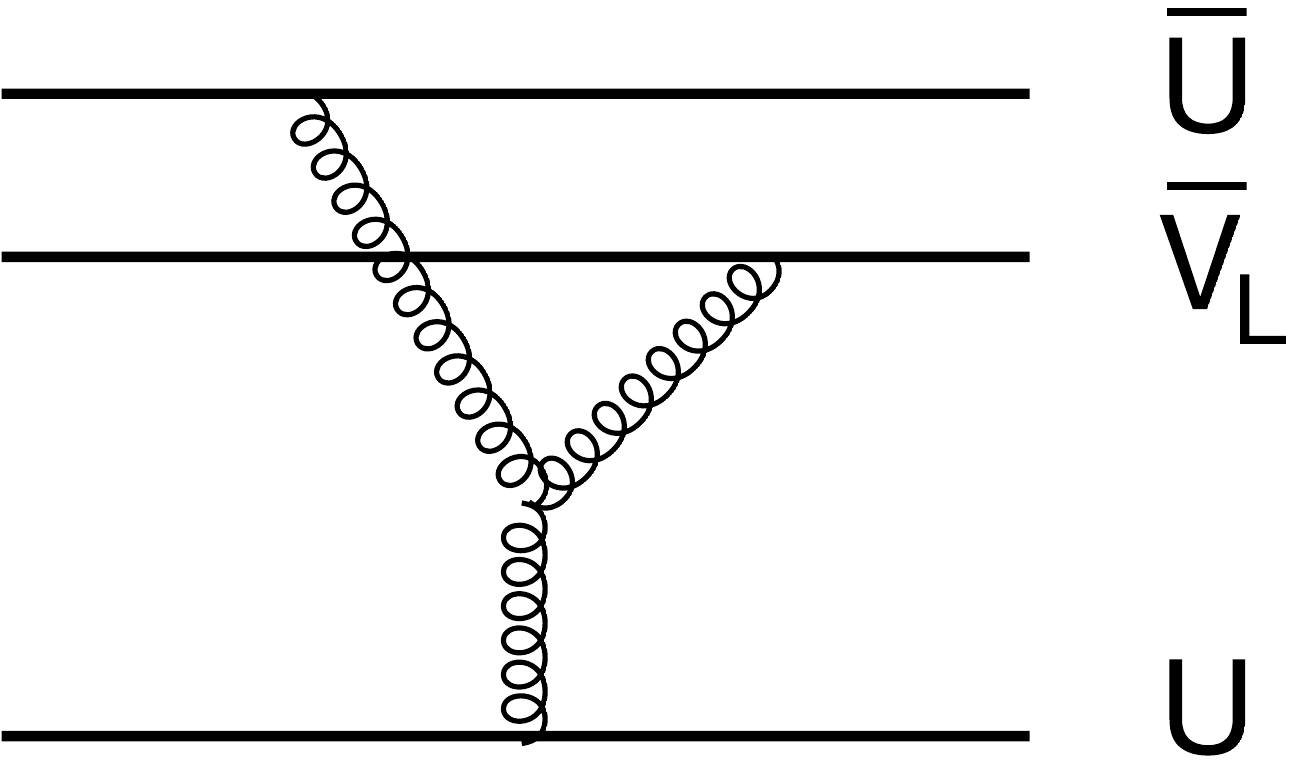}
\caption{An $\alpha_s$ suppressed correction to a correlator containing $\bar V_L$ which is negligible in the weak coupling limit.}
\label{7}
\end{figure}

As we have discussed above, the only difference between these two sets of diagrams 
 is that all the gluons exchanged between $\bar V_L$ and any given factor  $U$ connect to the left of any other gluons that might be exchanged by this $U$ and a different factor of $\bar U$ present in the amplitude. However any given $U$ can exchange at most two gluons. If these two gluons are exchanged between $U$ and $\bar V_L$, no further gluons are exchanged and the action of $\bar V_L$ is identical to the action of $\bar V$. If $U$ exchanges only one gluon with $\bar V_L$ and another gluon with some other factor of $\bar U$, it is still true that as far as elastic amplitude is concerned the action of $\bar V_L$ and $\bar V$ is identical. The difference only appears in the inelastic amplitude, but here again it appears as $\alpha_s$ suppressed correction through a diagram analogous to that of \fig{2}, see \fig{7}. This correction is not enhanced by the number of target gluons, and thus is indeed negligible in the dense-dilute limit. We therefore conclude that in the dense dilute limit we can safely replace $\bar V_L$ by $\bar V$. The same is obviously true for $\bar V_R$. Thus in the dense-dilute limit in $H_{RFT}$ we can replace
\begin{equation}\label{barvl}
\bar V_L\rightarrow \bar V;\ \ \ \ \ \ \bar V_R\rightarrow \bar V^\dagger
\end{equation}

Another simplification follows since any factor of $U$, $V_L$ or $V_R$ can be expanded to second order as only two gluons can be exchanged by any of the target gluons. Thus in the dense-dilute limit   we have 
\begin{equation}\label{vllim}
\begin{split}
&V_L(\mathbf{x}) = 1 + \int_{\mathbf{y}} ig\phi(\mathbf{x}-\mathbf{y}) t^e \mathcal{J}^e_{L}(\mathbf{y}) - \frac{g^2}{2}\int_{\mathbf{y},\mathbf{z}}\phi(\mathbf{x}-\mathbf{y}) \phi(\mathbf{x}-\mathbf{z}) t^et^d \mathcal{J}_L^e(\mathbf{y})\mathcal{J}_L^d(\mathbf{z})\, ,\\
&V_R(\mathbf{x}) = 1 - \int_{\mathbf{y}} ig\phi(\mathbf{x}-\mathbf{y}) t^e \mathcal{J}^e_{R}(\mathbf{y}) -\frac{g^2}{2}\int_{\mathbf{y},\mathbf{z}}\phi(\mathbf{x}-\mathbf{y}) \phi(\mathbf{x}-\mathbf{z}) t^et^d \mathcal{J}_R^e(\mathbf{y})\mathcal{J}_R^d(\mathbf{z})\, .\\
\end{split}
\end{equation}
With these simplification we now consider the RFT Hamiltonian. Let us concentrate on  $H^{(1)}_{RFT}$
\begin{equation}
H^{(1)}_{RFT} \approx\frac{1}{\pi g^2}\int_{\mathbf{x}} \bar{V}^{\beta\lambda}(\mathbf{x})\bar{V}^{\dagger \kappa\alpha}(\mathbf{x}) \partial^2\left[V_L^{\alpha\beta}(\mathbf{x})V_R^{ \lambda\kappa}(\mathbf{x})\right]\, .
\end{equation}
with the understanding that $V_L, V_R$ are expanded to second order.
The zeroth order in expansion, the product $V_LV_R$ is a constant and does not contribute to the Hamiltonian due to derivative acting on it. The first order also vanishes because it involves a factor $\mathrm{Tr} (t^a )=0$.  At second order there are three terms
\begin{equation}
\begin{split}
V_L^{\alpha\beta}(\mathbf{x})V_R^{\lambda\kappa}(\mathbf{x}) = \frac{g^2}{2}\int_{\mathbf{y},\mathbf{z}} i\phi(\mathbf{x}-\mathbf{y}) i\phi(\mathbf{x}-\mathbf{z})&\Big[-2 t^e_{\alpha\beta} t^d_{\lambda\kappa} \mathcal{J}^e_L(\mathbf{y})\mathcal{J}^d_R(\mathbf{z}) + (t^et^d)_{\alpha\beta}\delta^{\lambda\kappa} \mathcal{J}^e_L(\mathbf{y})\mathcal{J}^d_L(\mathbf{z})\\
& +(t^et^d)_{\lambda\kappa}\delta^{\alpha\beta} \mathcal{J}^e_R(\mathbf{y})\mathcal{J}^d_R(\mathbf{z})\Big].\\
\end{split}
\end{equation}
Substituting the above expression into $H_{RFT}$, one obtains
\begin{equation}\label{eq:twoEps_JIMWLK}
\begin{split}
H^{(1)}_{RFT}   &= \frac{1}{2\pi}\int_{\mathbf{x},\mathbf{y},\mathbf{z}}\partial^2_{\mathbf{x}}[i\phi(\mathbf{x}-\mathbf{y}) i\phi(\mathbf{x}-\mathbf{z})]\Big(-2\bar{V}^{\beta\lambda}(\mathbf{x})\bar{V}^{\dagger\kappa\alpha}(\mathbf{x})[t^e\mathcal{J}_L^e(\mathbf{y})]_{\alpha\beta}[t^d\mathcal{J}_R^d(\mathbf{z})]_{\lambda\kappa} \\
&+ [t^e\mathcal{J}^e_L(\mathbf{y})]_{\alpha\beta}[t^d\mathcal{J}^d_L(\mathbf{z}) ]_{\beta\alpha}+ [t^e\mathcal{J}^e_R(\mathbf{y})]_{\alpha\beta}[t^d\mathcal{J}^d_R(\mathbf{z}) ]_{\beta\alpha}\Big)\\
 =& \frac{1}{4\pi}\int_{\mathbf{x},\mathbf{y},\mathbf{z}}\partial^2_{\mathbf{x}}[i\phi(\mathbf{x}-\mathbf{y}) i\phi(\mathbf{x}-\mathbf{z})]\left[-2\bar{U}^{ed}(\mathbf{x})\mathcal{J}_L^e(\mathbf{y})\mathcal{J}_R^d(\mathbf{z}) + \mathcal{J}^e_L(\mathbf{y})\mathcal{J}^e_L(\mathbf{z}) + \mathcal{J}^e_R(\mathbf{y})\mathcal{J}^e_R(\mathbf{z}) \right]\\
=& \frac{1}{2\pi}\int_{\mathbf{x},\mathbf{y},\mathbf{z}}[i\partial_{\mathbf{x}}\phi(\mathbf{x}-\mathbf{y}) i\partial_{\mathbf{x}}\phi(\mathbf{x}-\mathbf{z})]\left[-2\bar{U}^{ed}(\mathbf{x})\mathcal{J}_L^e(\mathbf{y})\mathcal{J}_R^d(\mathbf{z}) + \mathcal{J}^e_L(\mathbf{y})\mathcal{J}^e_L(\mathbf{z}) + \mathcal{J}^e_R(\mathbf{y})\mathcal{J}^e_R(\mathbf{z}) \right].\\
\end{split}
\end{equation}
 Note that the spatial derivatives generate other terms 
 \begin{equation}
 \partial^2_{\mathbf{x}}  (\phi(\mathbf{x}-\mathbf{y}) \phi(\mathbf{x}-\mathbf{z})) = [\partial_{\mathbf{x}}^2\phi(\mathbf{x}-\mathbf{y})] \phi(\mathbf{x}-\mathbf{z}) + \phi(\mathbf{x}-\mathbf{y}) [\partial^2_{\mathbf{x}}\phi(\mathbf{x}-\mathbf{z})] +2\partial_{\mathbf{x}}\phi(\mathbf{x}-\mathbf{y})\partial_{\mathbf{x}} \phi(\mathbf{x}-\mathbf{z}).
 \end{equation}
 However, with $\partial_{\mathbf{x}}^2\phi(\mathbf{x}-\mathbf{y}) = g\delta(\mathbf{x}-\mathbf{y})$ and $\partial^2_{\mathbf{x}}\phi(\mathbf{x}-\mathbf{z}) = g\delta(\mathbf{x}-\mathbf{z})$, performing the integration over $\mathbf{x}$ and using the relations $\bar{U}^{ed}(\mathbf{y})\mathcal{J}^e_L(\mathbf{y}) = \mathcal{J}^d_R(\mathbf{y})$ and $\bar{U}^{ed}(\mathbf{z})\mathcal{J}^d_R(\mathbf{z}) = \mathcal{J}^e_L(\mathbf{z}) $, these addtional terms cancel each other. Thus only the term where the two derivatives separately act on $\phi(\mathbf{x}-\mathbf{y})$ and $\phi(\mathbf{x}-\mathbf{z})$ survives.
 Performing the same calculation for $H^{(1)c}_{RFT}$ we find to this order an identical result. Thus in the dense-dilute approximation we get
 \begin{eqnarray}
 H_{RFT}&\rightarrow& H_{JIMWLK} \\
 &=&\frac{1}{2\pi}\int_{\mathbf{x},\mathbf{y},\mathbf{z}}[i\partial_{\mathbf{x}}\phi(\mathbf{x}-\mathbf{y}) i\partial_{\mathbf{x}}\phi(\mathbf{x}-\mathbf{z})]\left[-2\bar{U}^{ed}(\mathbf{x})\mathcal{J}_L^e(\mathbf{y})\mathcal{J}_R^d(\mathbf{z}) + \mathcal{J}^e_L(\mathbf{y})\mathcal{J}^e_L(\mathbf{z}) + \mathcal{J}^e_R(\mathbf{y})\mathcal{J}^e_R(\mathbf{z}) \right].\nonumber
\end{eqnarray}

There is one subtlety in this derivation which we need to address, i.e. at what order does the correction to \eq{barvl} affect the calculation. To answer this we need to develop a controlled expansion of $H_{RFT}$ in the dense-dilute limit.
To do this
we note that although we have justified Eqs.(\ref{barvl}) and (\ref{vllim}) by analyzing the contributions to the S-matrix generated by exchanges of at most two gluon, the same result can be obtained formally by taking the limit of small $\rho$. It is obvious that at small $\rho$, the operators $V_L$ and $V_R$ should be simply expanded in power series in $\mathcal{J}_{L(R)}$ to the leading order to which the Hamiltonian does not vanish, leading to \eq{vllim}. On the other hand at small $\rho$ we should also expand $\mathcal{I}_{L(R)}$ to leading order in $\rho$, which gives
\begin{equation}\label{irho}
\mathcal{I}^a_L=\mathcal{I}^a_R=\frac{-i}{g}\frac{\delta}{\delta\alpha^a(\mathbf{x})};\ \ \ \ \ \ \ \ \ \bar V_L=V; \ \ \ \bar V_R=\bar V^\dagger.
\end{equation}
In fact expansion  in powers of $\rho$ is the proper formal way to derive the form of the Hamiltonian in the dense-dilute limit.

Formally expanding $H^{(1)}_{RFT}$ in powers of $\rho$ we see that $H_{JIMWLK}$ arises at order $\rho^2$ by multiplying the $O(1)$ term in $\bar V_L\bar V_R$ and $O(\rho^2)$ term in $V_LV_R$. However we also have to consider a possible contribution arising from $O(\rho)$ term in $\bar V_L\bar V_R$ (the first order correction to \eq{irho}) multiplied by $O(\rho)$ term in $V_LV_R$. 
We write this additional term as 
\begin{equation}\label{eq:linear_term_v2}
\begin{split}
\delta H_{JIMWLK}= & \frac{1}{\pi g^2}\int_{\mathbf{x}}\left( \bar{V}_L^{\beta\lambda}(\mathbf{x})\bar{V}^{\dagger \kappa\alpha}  + \bar{V}^{\beta\lambda} \bar{V}_R^{ \kappa\alpha}(\mathbf{x})\right)\partial^2\left[V_L^{\alpha\beta}(\mathbf{x})\delta^{\lambda\kappa}+ \delta^{\alpha\beta}V_R^{ \lambda\kappa}(\mathbf{x})\right]\,\\
=&\frac{1}{\pi g^2}\int_{\mathbf{x}}\left( (\bar{V}_L\bar{V}^{\dagger})^{\beta\alpha}+ (\bar{V} \bar{V}_R)^{\beta\alpha}\right)\left[ig^2t^e_{\alpha\beta}\mathcal{J}_L^e(\mathbf{x})\right]+\left( (\bar{V}^{\dagger}\bar{V}_L)^{\kappa\lambda}+ (\bar{V}_R \bar{V})^{\kappa\lambda}\right)\left[-ig^2t^d_{\lambda\kappa}\mathcal{J}_R^d(\mathbf{x})\right]\,\\
\end{split}
\end{equation}
Here $\bar V_L$ and $\bar V_R$ are understood as expanded to $O(g)$, however we will not need the explicit from of this expansion, since we will show that this expression vanishes.

We use the two identities
\begin{equation}
\bar{V}^{\dagger}_{\lambda \alpha} t^e_{\alpha\beta} \bar{V}_{\beta\gamma}  = \bar{U}^{ed} t^d_{\lambda\gamma} 
\end{equation}
and  
\begin{equation}
\bar{U}^{\dagger de}(\mathbf{x})\mathcal{J}_L^e(\mathbf{x}) = \mathcal{J}_R^d(\mathbf{x}).
\end{equation}
Here, as before  $\bar{V}^{\dagger} = \mathrm{exp}\{-t^a\frac{\delta}{\delta\rho^a}\}$ and $\bar{V} = \mathrm{exp}\{t^a\frac{\delta}{\delta\rho^a}\}$ are defined in the fundamental representation while $\bar{U} = \mathrm{exp}\{T^a \frac{\delta}{\delta \rho^a}\}$ is defined in the adjoint representation. 
We then calculate
\begin{equation}
(\bar{V}_L\bar{V}^{\dagger})^{\beta\alpha} t^e_{\alpha\beta} \mathcal{J}_L^e =  \bar{V}_L^{\beta\lambda} \left[\bar{U}^{ed} t^d_{\lambda\gamma} \bar{V}^{\dagger \gamma\beta}\right] \mathcal{J}^e_L = \mathcal{J}_R^d t^d_{\lambda\gamma} (\bar{V}^{\dagger} \bar{V}_L)^{\gamma\lambda}
\end{equation}
and 
\begin{equation}
(\bar{V} \bar{V}_R)^{\beta\alpha}t^e_{\alpha\beta}\mathcal{J}_L^e =\left[ \bar{U}^{ed} \bar{V}_{\alpha\lambda} t^d_{\lambda\gamma}\right] \bar{V}_R^{ \gamma\alpha} \mathcal{J}_L^e = \mathcal{J}^d_R t^d_{\lambda\gamma} (\bar{V}_R \bar{V})^{\gamma\lambda}.
\end{equation}
Thus the four terms  in \eq{eq:linear_term_v2} pairwise cancel. 

We have thus proved that when expanded to second order in $\rho$, the Hamiltonian $H^{(1)}_{RFT}$ reproduces $H_{JIMLWK}$. It is obvious that the same is true for $H^{(1)c}_{RFT}$, since $H_{JIMWLK}$ is charge conjugation invariant.

If instead of expanding in powers of $\rho$, we expand in powers of $\delta/\delta\rho$, the leading order expansion gives $H_{KLWMIJ}$, \eq{klwmij}. This is easily done explicitly, but the final result is obvious by duality.

Finally we note that the exact same result is obtained if we were to use the left and right Wilson lines not in the fundamental but in any other representation of $SU(N)$.   The only property of the $SU(N)$ matrices that is needed to derive $H_{JIMWLK}$ in \eq{eq:twoEps_JIMWLK} is
\begin{equation}
{\rm Tr}[\bar V^\dagger t^e\bar Vt^d]=\frac{1}{2}\bar U^{ed}
\end{equation}
for a fundamental matrix $\bar V$ and an adjoint matrix $\bar U$. However a similar relation holds for $SU(N)$ matrices in any representation $D$
\begin{equation}
{\rm Tr}[\bar U^\dagger_DT^a_D\bar U_D T^b_D]=\frac{C_2(D)R_D}{N^2-1}\bar U^{ab}.
\end{equation}
Here $\bar U_D$ is a matrix and $T_D$ is a generator in an arbitrary representation $D$ of $SU(N)$,  and $C_2(D)$ and $R_D$ are the second Casimir and the dimensionality of $D$ respectively.  Thus using $U_D$ and $\bar U_D$ in any representation in the definition of $H_{RFT}$ will reproduce $H_{JIMWLK}$ and $H_{KLWMIJ}$  in expansion once the overall normalization is adjusted.

\section{Continuous symmetries}
Let us now discuss the continuous symmetries of $H_{RFT}$.  As we have mentioned above, both the JIMWLK and the KLWMIJ Hamiltonians have a continuous $SU(N)\times SU(N)$ symmetry, albeit those are distinct symmetry transformations. The $SU(N)\times SU(N)$ symmetry of $H_{JIMWLK}$ is generated by the charges
\begin{equation}
Q^a_L=\int d^2\mathbf{z}\ \mathcal{J}_L^a(\mathbf{z}); \ \ \ \  Q^a_R=\int d^2\mathbf{z}\ \mathcal{J}_R^a(\mathbf{z})
\end{equation}
while the $SU(N)\times SU(N)$ symmtery of $H_{KLWMIJ}$ by
\begin{equation}
\bar Q^a_L=\int d^2\mathbf{z}\ \mathcal{I}_L^a(\mathbf{z}); \ \ \ \  \bar Q^a_R=\int d^2\mathbf{z}\ \mathcal{I}_R^a(\mathbf{z}).
\end{equation}

It is an interesting question which of these symmetries are also the symmetries of the self dual $H_{RFT}$ \eq{eq:correct_RFT}. The question is not completely straightforward to answer even though we do have an explicit representation of the charge operators on the RFT Hilbert space. The reason is that  the commutation relations between $\mathcal{J}_{L(R)}$ and $\bar V_{L(R)}$ as well as between $\mathcal{I}_{L(R)}$ and $V_{L(R)}$ are quite complicated.
 We will nevertheless try to answer this question, using a perturbative expansion. Our answer is somewhat surprising: the symmetry of $H_{RFT}$ appears to be $SU(N)\times SU(N)\times SU(N)$\footnote{To be precise, while  $SU(N)\times SU(N)$ is there, the third  $SU(N)$  does not necessary form  a direct product with the first two. We have not 
 attempted to write down the full algebra of the currents, which appears to  be quite complicated.}.
 
 We start with discussing the vector part of the group, which is the easiest and can be analyzed without recourse to perturbation theory. 
 
 To better organize the calculation, we rescale the charge density  $\tilde{\rho}^a(\mathbf{x}) = g \rho^a(\mathbf{x})$ and also introduce $\tilde{\phi}(\mathbf{x}-\mathbf{y})=\frac{1}{g}\phi(\mathbf{x}-\mathbf{y})$. Then $\mathcal{J}_L^a, \mathcal{J}_R^a, \mathcal{I}_L^a,  \mathcal{I}_R^a$ can be Taylor expanded by counting the powers of the coupling constant $g$. We will use this expansion in this and the next sections. We will refer to this counting in powers of the coupling constant as the "BFKL counting", since it is equivalent to simultaneous expansion in powers of $\rho$ and $\delta/\delta\rho$.

\subsection{ The vector $SU_V(N)$  symmetry}

The analysis of the vector symmetry is facilitated by the following simple observation
\begin{equation}\label{equal}
Q_L-Q_R=\bar Q_R-\bar Q_L.
\end{equation}
To prove this we note that
\begin{equation}
\mathcal{J}^a_L(\mathbf{z})-\mathcal{J}^a_R(\mathbf{z})=\tilde\rho^b(\mathbf{z})T^a_{bc}\frac{\delta}{\delta\tilde\rho^c(\mathbf{z})}
\end{equation}
\begin{equation}
\mathcal{I}^a_R(\mathbf{z})-\mathcal{I}^a_L(\mathbf{z})=\tilde\alpha^b(\mathbf{z})T^a_{bc}\frac{\delta}{\delta\tilde\alpha^c(\mathbf{z})}=\int d^2\mathbf{y}\tilde\phi(\mathbf{z}-\mathbf{y})\tilde\rho^b(\mathbf{y})T^a_{bc}\partial^2_{\mathbf{z}}\frac{\delta}{\delta\tilde\rho^c(\mathbf{z})}.
\end{equation}
Integrating by parts we find
\begin{equation}
\bar Q_R-\bar Q_L=\int d^2\mathbf{z}d^2\mathbf{y}\tilde\phi(\mathbf{z}-\mathbf{y})\tilde\rho^b(\mathbf{y})T^a_{bc}\partial^2_{\mathbf{z}}\frac{\delta}{\delta\tilde\rho^c(\mathbf{z})}=\int d^2\mathbf{z}\tilde\rho^b(\mathbf{z})T^a_{bc}\frac{\delta}{\delta\tilde\rho^c(\mathbf{z})}=Q_L-Q_R.
\end{equation}
It is now straightforward to check that the vector $SU_V(N)$ transformation generated by $Q_L-Q_R$ is the symmetry of $H_{RFT}$. 
By virtue of \eq{equal} the charge $Q_V^a\equiv Q^a_L-Q^a_R$ acts as a rotation generator on all the currents, i.e.
\begin{equation}
\begin{split}
&\left[Q_V^a, \mathcal{J}_L^b\right] = if^{abc}\mathcal{J}_L^c\, ,\\
&\left[Q_V^a, \mathcal{J}_R^b\right] = if^{abc}\mathcal{J}_R^c\, .\\
&\left[Q_V^a, \mathcal{I}_L^b\right] = if^{abc}\mathcal{I}_L^c\, ,\\
&\left[Q_V^a, \mathcal{I}_R^b\right] = if^{abc}\mathcal{I}_R^c\, .\\
\end{split}
\end{equation}
It then follows that for a finite group transformation 
\begin{equation}
\hat{W} = \mathrm{exp}\left\{i\lambda^a Q_V^a\right\}
\end{equation}
we have
\begin{equation}
\hat{W}^{\dagger} \mathcal{J}_{L(R)}^a(\mathbf{x}) \hat{W} = \mathcal{W}_A^{ab}\mathcal{J}_{L(R)}^b(\mathbf{x}); \ \ \hat{W}^{\dagger} \mathcal{I}_{L(R)}^a(\mathbf{x}) \hat{W} = \mathcal{W}_A^{ab}\mathcal{I}_{L(R)}^b(\mathbf{x});
\end{equation}
with 
\begin{equation}
\mathcal{W}_A^{ab} = \left[e^{i\lambda^dT^d}\right]^{ab}.
\end{equation}
As a consequence
\begin{equation}\label{eq:VL_tranformation}
\begin{split}
&\hat{W}^{\dagger} V_{L}^{\alpha\beta}(\mathbf{x}) \hat{W} 
=\left(\mathcal{W}_F V_L(\mathbf{x}) \mathcal{W}_F^{\dagger}\right)^{\alpha\beta}
\end{split}
\end{equation}
with the fundamental representation matrix
\begin{equation}
\mathcal{W}_F^{\kappa\beta} = \left[e^{i\lambda^e t^e}\right]^{\kappa\beta}.
\end{equation}
The same transformation as in \eq{eq:VL_tranformation} applies to $V_R(\mathbf{x})$, as well as to $\bar V_{L(R)}$.
It is now obvious that $H_{RFT}$ is invariant under $SU_V(N)$.

\subsection{Is $SU_L(N)$ there?}
Let us now consider other transformations generated by the left and right charges. The analysis for all of them is similar, and we will concentrate on $Q_L$. The question we are asking, does $Q_L$ commute with $H_{RFT}$?

What is the action of $Q^a_L$ on the building blocks of $H_{RFT}$? The answer for $V_L$ and $V_R$ is obvious. Under the $SU_L(N)$ transfromation
\begin{equation}
\hat{S} = \mathrm{exp}\left\{i \lambda^aQ_L^a\right\} 
\end{equation}
we have 
\begin{equation}
\hat{S}^{\dagger} \mathcal{J}_L^e(\mathbf{z}) \hat{S}= \mathcal{S}^{ed}_A\mathcal{J}_L^d(\mathbf{z}); \ \ \ \ \ \ \ \hat{S}^{\dagger} \mathcal{J}_R^e(\mathbf{z}) \hat{S}=\mathcal{J}_R^e(\mathbf{z}) 
\end{equation}
with 
\begin{equation}
\mathcal{S}^{ed}_A(\mathbf{x}) = \left[e^{i\lambda^a T^a} \right]^{ed}.
\end{equation}
As a consequence, 
\begin{equation}
\begin{split}
&\hat{S}^{\dagger} V_L^{\alpha\beta}(\mathbf{x}) \hat{S}  = \mathcal{S}_F^{\alpha\gamma} V_L^{\gamma\kappa}(\mathbf{x}) \mathcal{S}_F^{\dagger\kappa\beta}; \ \ \ \ \ \ \ \hat{S}^{\dagger} V_R^{\alpha\beta}(\mathbf{x}) \hat{S}  =  V_R^{\alpha\beta}(\mathbf{x}) 
\end{split}
\end{equation}
with 
\begin{equation}
S_F^{\kappa\beta} = \left[e^{i\lambda^e t^e}\right]^{\kappa\beta}.
\end{equation}

What is the transformation of $\bar V_L$ and $\bar V_R$?
Examining the expression for $H_{RFT}$ we see that if the transformation was 
\begin{equation}\label{eq:barVL_transform}
\hat{S}^{\dagger} \bar{V}_L^{\beta\gamma}(\mathbf{x}) \hat{S} = \mathcal{S}_F^{\beta\kappa} \bar{V}_L^{\kappa\gamma}(\mathbf{x});  \ \ \ \ \ \ \hat{S}^{\dagger} \bar{V}_R^{\beta\gamma}(\mathbf{x}) \hat{S} = \bar{V}_R^{\beta \kappa}\mathcal{S}_F^{\kappa\gamma} (\mathbf{x}); \ \ \ \ \ \ \ \ \ \ \  ({\rm true ~ or ~ false~?})
\end{equation}
the Hamiltonian would be invariant under $SU_L(N)$. Indeed if instead of $\bar V_L$ and $\bar V_R$ we had $\bar V$ and $\bar V^\dagger$, this would be the case. This is precisely what happens in the JIMWLK limit.

The transformation \eq{eq:barVL_transform} is equivalent to the commutation relation
\begin{equation}\label{eq:JL_VL_commutator}
\left[Q_L^a, \bar{V}_L^{\alpha\beta}(\mathbf{x})\right] = -\left(t^a\bar{V}_L(\mathbf{x})\right)^{\alpha\beta} \ \ \ \ \ \ \ \ ({\rm true ~ or ~ false~?})
\end{equation}
and similarly for $\bar V_R$.

We were unable to  calculatie the commutation relation in  \eq{eq:JL_VL_commutator} in a closed form. However we were able to calculate first several orders in perturbative expansion in $g$. We performed the calculation in the BFKL counting of orders of $g$. The details of the calculation are presented in the Appendix  A. Our results are the following.

We have calculated  the commutator between $Q^a_L$ and $\bar V_L$   up to order $g^3$ and found that relation \eq{eq:JL_VL_commutator} holds up to order $g^2$, but is violated at order $g^3$.

We have also calculated the commutator of $Q_L^a$ with the Hamiltonian $[Q_L^a,H_{RFT}]$ up to order $g^6$. We have found that this commutator vanishes up to this order. This leads us to believe that even though \eq{eq:JL_VL_commutator} is not satisfied, the $SU_L(N)$ is indeed a symmetry of $H_{RFT}$. We stress that  we do not have a closed form proof of this, but only perturbative calculation to order $g^6$.

The analysis of $\bar Q^a_L$ is identical, since $Q$ and $\bar Q$ are related by duality transformation. Thus we believe that $\bar Q^a_L$ also commutes with the Hamiltonian.

If this is indeed the case, the continuous symmetry of $H_{RFT}$ is at least $SU(N)\times SU(N)\times SU(N)$. In fact the symmetry could be even larger since we have not calculated the commutators $[Q_L,\bar Q_L]$. If this commutator does not close on any of the four charges (or their products) $Q_{L(R)}$, $\bar Q_{L(R)}$ the symmetry group is larger. We have not investigated this question any further.

\section{Is this the ``Diamond action"?}
The family of Hamiltonians that we have identified carries uncanny resemblance to the so called "Diamond action" suggested in \cite{diamond} and also discussed in \cite{Balitsky05}.
There is of course a host of differences between our approach and that of \cite{diamond} and \cite{Balitsky05}. On the technical level we are dealing with the Hamiltonian formulation of RFT together with the accompanying field algebra and the structure of the RFT Hilbert space, while these references strive to derive the effective action in terms of certain Wilson line functions. On the other hand \cite{diamond} and \cite{Balitsky05}  derive the action directly from QCD (although in both cases certain not entirely straightforward approximations are utilized)  whereas our expression is an ansatz constrained by the expected symmetries and the appropriate limiting forms.

Nevertheless, abstracting ourselves from these differences we can compare $H_{RFT}$ with the effective action of \cite{diamond}. We concentrate on the Hamiltonian \eq{eq:correct_RFT} defined with Wilson line in the \textit{adjoint representation}.
\begin{equation}
\begin{split}
&U_L(\mathbf{x}) = \mathrm{Exp}\left\{i\int_{\mathbf{y}} g\phi(\mathbf{x}-\mathbf{y}) T^e \mathcal{J}^e_L(\mathbf{y})\right\}\\
&U_R(\mathbf{x}) =\mathrm{Exp}\left\{-i\int_{\mathbf{y}}g \phi(\mathbf{x}-\mathbf{y}) T^e \mathcal{J}^e_R(\mathbf{y})\right\}\\
&\bar{U}_L(\mathbf{x}) = \mathrm{Exp}\left\{i\int_{\mathbf{y}} g\phi(\mathbf{x}-\mathbf{y}) T^e \mathcal{I}^e_L(\mathbf{y})\right\}\\
&\bar{U}_R(\mathbf{x}) = \mathrm{Exp}\left\{-i\int_{\mathbf{y}} g\phi(\mathbf{x}-\mathbf{y}) T^e \mathcal{I}^e_R(\mathbf{y})\right\}.\\
\end{split}
\end{equation}
 In this case the two terms in \eq{eq:correct_RFT} are equal and we have
\begin{equation}\label{ha}
H^A_{RFT}=\frac{1}{2\pi g^2N}\int d^2\mathbf{x} \, \partial^2[\bar{U}_L^{bc}(\mathbf{x}) \bar{ U}^{ da}_R(\mathbf{x})] U_L^{ab}(\mathbf{x}) U_{R}^{ cd}(\mathbf{x})\,. 
\end{equation}
It is easily checked that with the correspondence
\begin{equation}
U_L\rightarrow W_{-\infty}, \,\,U_R \rightarrow W_{\infty}^{\dagger}, \,\, \bar{U}_L \rightarrow V_{-\infty}, \, \, \bar{U}_R \rightarrow V_{\infty}^{\dagger} 
\end{equation}
our \eq{ha} looks identical to the effective action suggested in \cite{diamond}. However beyond the looks there are significant differences between the two. In particular in \cite{diamond} the four Wilson lines are not independent, but satisfy the so called diamond condition
\begin{equation}\label{diamondc}
V^\dagger_{\infty}W_{-\infty} V_{-\infty}W^\dagger_\infty=1\,.
\end{equation}
This relation was essential in the derivation of \cite{diamond} and only using this relation the effective action obtained in \cite{diamond} could be written in the form \eq{ha}.
On the other hand in our framework, although all four Wilson line operators are expressible in terms of $\rho$ and $\delta/\delta\rho$,  there is no such condition that constrains the four.

We  can  check \eq{diamondc}  explicitly, expanding all the operators $U_{L,R}$ and $ \bar U_{L,R}$ to first order in the respective left and right charge densities.
In our notations \eq{diamondc})  corresponds to 
\begin{equation}\label{diamondus}
\bar{U}_R(\mathbf{x}) U_L(\mathbf{x}) \bar{U}_L(\mathbf{x}) U_R(\mathbf{x}) =1\,.
\end{equation}

 We will calculate the LHS of \eq{diamondus}  to second order in $g$. To this order we need
\begin{equation}
\begin{split}
\mathcal{J}_L^a(\mathbf{x})  = &\frac{1}{g}\left[\frac{1}{2}gT^e\frac{\delta}{\delta \tilde{\rho}^e(\mathbf{x})} \left(\coth{\left[\frac{1}{2}gT^e\frac{\delta}{\delta \tilde{\rho}^e(\mathbf{x})}\right]}-1\right)\right]^{ba}\tilde{\rho}^b(\mathbf{x}) \, ,\\
=&\frac{1}{g}\tilde{\rho}^a(\mathbf{x}) - \frac{1}{2} \tilde{\rho}^b(\mathbf{x})T^e_{ba} \frac{\delta}{\delta \tilde{\rho}^e(\mathbf{x})}  +\mathcal{O}(g)\\
\end{split}
\end{equation}
\begin{equation}
\begin{split}
\mathcal{J}_R^a(\mathbf{x})  =& \frac{1}{g}
 \left[\frac{1}{2}gT^e\frac{\delta}{\delta \tilde{\rho}^e(\mathbf{x})} \left(\coth{\left[\frac{1}{2}gT^e\frac{\delta}{\delta \tilde{\rho}^e(\mathbf{x})}\right]}+1\right)\right]^{ba}\tilde{\rho}^b(\mathbf{x}) \, ,\\
=&\frac{1}{g}\tilde{\rho}^a(\mathbf{x}) + \frac{1}{2} \tilde{\rho}^b(\mathbf{x})T^e_{ba} \frac{\delta}{\delta \tilde{\rho}^e(\mathbf{x})}  +\mathcal{O}(g).\\
\end{split}
\end{equation}

\begin{equation}
\begin{split}
\mathcal{I}_L^a(\mathbf{x}) & = \frac{-i}{g}\partial^2 \frac{\delta}{\delta \tilde{\rho}^b(\mathbf{x})}  \left[\frac{1}{2}igT^e\frac{1}{\partial^2} \tilde{\rho}^e(\mathbf{x})\left(\coth{\left[\frac{1}{2}igT^e\frac{1}{\partial^2} \tilde{\rho}^e(\mathbf{x})\right]}-1\right)\right]^{ba}\, ,\\
&=\frac{-i}{g}\partial^2 \frac{\delta}{\delta \tilde{\rho}^a(\mathbf{x})} - \frac{1}{2} T^e_{ba}\partial^2 \frac{\delta}{\delta \tilde{\rho}^b(\mathbf{x})} \int_{\mathbf{z}} \tilde{\phi}(\mathbf{x}-\mathbf{z})\tilde{\rho}^e(\mathbf{z}) + \mathcal{O}(g).
\end{split}
\end{equation}
\begin{equation}
\begin{split}
\mathcal{I}_R^a(\mathbf{x})  &= \frac{-i}{g^2}\partial^2 \frac{\delta}{\delta \rho^b(\mathbf{x})}  \left[\frac{1}{2}T^eig^2\frac{1}{\partial^2} \rho^e(\mathbf{x})\left(\coth{\left[\frac{1}{2}T^eig^2\frac{1}{\partial^2} \rho^e(\mathbf{x})\right]}+1\right)\right]^{ba}\, ,\\
&=\frac{-i}{g}\partial^2 \frac{\delta}{\delta \tilde{\rho}^a(\mathbf{x})} +\frac{1}{2} T^e_{ba}\partial^2 \frac{\delta}{\delta \tilde{\rho}^b(\mathbf{x})} \int_{\mathbf{z}} \tilde{\phi}(\mathbf{x}-\mathbf{z})\tilde{\rho}^e(\mathbf{z}) + \mathcal{O}(g).
\end{split}
\end{equation}

From the definition of $\mathcal{I}_L^a$ and $\mathcal{I}_R^a$, one obtains
\begin{equation}
\begin{split}
&ig^2\int_{\mathbf{y}} \tilde{\phi}(\mathbf{x}-\mathbf{y}) T^e\mathcal{I}_L^e(\mathbf{y})\\
=&ig^2\int_{\mathbf{y}} \tilde{\phi}(\mathbf{x}-\mathbf{y}) T^e \left( \frac{-i}{g}\partial^2_{\mathbf{y}} \frac{\delta}{\delta \tilde{\rho}^e(\mathbf{y})}  - \frac{1}{2}\partial^2_{\mathbf{y}} \frac{\delta}{\delta \tilde{\rho}^b(\mathbf{y})}  T^a_{be} \int_{\mathbf{z}}\tilde{\phi}(\mathbf{y}-\mathbf{z}) \tilde{\rho}^a(\mathbf{z})+\mathcal{O}(g)\right)\\
=&gT^e \frac{\delta}{\delta \tilde{\rho}^e(\mathbf{x})} - \frac{1}{2}ig^2 T^eT^e_{ab} \int_{\mathbf{y},\mathbf{z}} \tilde{\phi}(\mathbf{x}-\mathbf{y})\tilde{\phi}(\mathbf{y}-\mathbf{z})\partial^2_{\mathbf{y}} \frac{\delta}{\delta \tilde{\rho}^b(\mathbf{y})} \tilde{\rho}^a(\mathbf{z}) + \mathcal{O}(g^3)\\
=&gT^e \frac{\delta}{\delta \tilde{\rho}^e(\mathbf{x})} - \frac{1}{2}ig^2 T^eT^e_{ab} \int_{\mathbf{y}} \tilde{\phi}(\mathbf{x}-\mathbf{y})\frac{\delta}{\delta \tilde{\rho}^b(\mathbf{y})} \tilde{\rho}^a(\mathbf{y}) - \frac{1}{2}ig^2 T^eT^e_{ab} \int_{\mathbf{z}} \tilde{\phi}(\mathbf{x}-\mathbf{z}) \frac{\delta}{\delta \tilde{\rho}^b(\mathbf{x})} \tilde{\rho}^a(\mathbf{z})\\
&- ig^2 T^eT^e_{ab} \int_{\mathbf{y},\mathbf{z}} \partial_{\mathbf{y}}\tilde{\phi}(\mathbf{x}-\mathbf{y})\partial_{\mathbf{y}}\tilde{\phi}(\mathbf{y}-\mathbf{z}) \frac{\delta}{\delta \tilde{\rho}^b(\mathbf{y})} \tilde{\rho}^a(\mathbf{z}) + \mathcal{O}(g^3).
\end{split}
\end{equation}
We have used integration by parts. As a consequence
\begin{equation}
\begin{split}
&\bar{U}_L(\mathbf{x}) = \mathrm{exp}\left\{ig^2\int_{\mathbf{y}} \tilde{\phi}(\mathbf{x}-\mathbf{y}) T^e\mathcal{I}_L^e(\mathbf{y})\right\}\\
=&1+gT^e \frac{\delta}{\delta \tilde{\rho}^e(\mathbf{x})} - \frac{1}{2}ig^2 T^eT^e_{ab} \int_{\mathbf{y}} \tilde{\phi}(\mathbf{x}-\mathbf{y})\frac{\delta}{\delta \tilde{\rho}^b(\mathbf{y})} \tilde{\rho}^a(\mathbf{y}) - \frac{1}{2}ig^2 T^eT^e_{ab} \int_{\mathbf{z}} \tilde{\phi}(\mathbf{x}-\mathbf{z}) \frac{\delta}{\delta \tilde{\rho}^b(\mathbf{x})} \tilde{\rho}^a(\mathbf{z})\\
&- ig^2 T^eT^e_{ab} \int_{\mathbf{y},\mathbf{z}} \partial_{\mathbf{y}}\tilde{\phi}(\mathbf{x}-\mathbf{y})\partial_{\mathbf{y}}\tilde{\phi}(\mathbf{y}-\mathbf{z}) \frac{\delta}{\delta \tilde{\rho}^b(\mathbf{y})} \tilde{\rho}^a(\mathbf{z}) + \frac{1}{2!} g^2T^eT^d \frac{\delta}{\delta \tilde{\rho}^e(\mathbf{x})} \frac{\delta}{\delta \tilde{\rho}^d(\mathbf{x})}+ \mathcal{O}(g^3)\\
\end{split}
\end{equation}
and 
\begin{equation}
\begin{split}
&\bar{U}_R(\mathbf{x}) = \mathrm{exp}\left\{-ig^2\int_{\mathbf{y}} \tilde{\phi}(\mathbf{x}-\mathbf{y}) T^e\mathcal{I}_R^e(\mathbf{y})\right\}\\
=&1-gT^e \frac{\delta}{\delta \tilde{\rho}^e(\mathbf{x})} - \frac{1}{2}ig^2 T^eT^e_{ab} \int_{\mathbf{y}} \tilde{\phi}(\mathbf{x}-\mathbf{y})\frac{\delta}{\delta \tilde{\rho}^b(\mathbf{y})} \tilde{\rho}^a(\mathbf{y}) - \frac{1}{2}ig^2 T^eT^e_{ab} \int_{\mathbf{z}} \tilde{\phi}(\mathbf{x}-\mathbf{z}) \frac{\delta}{\delta \tilde{\rho}^b(\mathbf{x})} \tilde{\rho}^a(\mathbf{z})\\
&- ig^2 T^eT^e_{ab} \int_{\mathbf{y},\mathbf{z}} \partial_{\mathbf{y}}\tilde{\phi}(\mathbf{x}-\mathbf{y})\partial_{\mathbf{y}}\tilde{\phi}(\mathbf{y}-\mathbf{z}) \frac{\delta}{\delta \tilde{\rho}^b(\mathbf{y})} \tilde{\rho}^a(\mathbf{z}) + \frac{1}{2!} g^2T^eT^d \frac{\delta}{\delta \tilde{\rho}^e(\mathbf{x})} \frac{\delta}{\delta \tilde{\rho}^d(\mathbf{x})}+ \mathcal{O}(g^3).\\
\end{split}
\end{equation}
On the other hand, from 
\begin{equation}
\begin{split}
&ig^2\int_{\mathbf{y}} \tilde{\phi}(\mathbf{x}-\mathbf{y}) T^e \mathcal{J}_L^e(\mathbf{y})\\
=&ig^2\int_{\mathbf{y}} \tilde{\phi}(\mathbf{x}-\mathbf{y}) T^e \left(\frac{1}{g}\tilde{\rho}^e(\mathbf{y}) - \frac{1}{2} \tilde{\rho}^b(\mathbf{y}) T^a_{be} \frac{\delta}{\tilde{\rho}(\mathbf{y})} + \mathcal{O}(g^3)\right)\\
=& ig\int_{\mathbf{y}} \tilde{\phi}(\mathbf{x}-\mathbf{y})T^e\tilde{\rho}^e(\mathbf{y}) - \frac{1}{2}ig^2 T^eT^e_{ab}\int_{\mathbf{y}}\tilde{\phi}(\mathbf{x}-\mathbf{y})\tilde{\rho}^b(\mathbf{y}) \frac{\delta}{\tilde{\rho}^a(\mathbf{y})} + \mathcal{O}(g^3).
\end{split}
\end{equation}
one obtains
\begin{equation}
\begin{split}
&U_L(\mathbf{x}) = \mathrm{exp}\left\{ig^2\int_{\mathbf{y}} \tilde{\phi}(\mathbf{x}-\mathbf{y}) T^e \mathcal{J}_L^e(\mathbf{y})\right\}\\
=&1+ ig\int_{\mathbf{y}} \tilde{\phi}(\mathbf{x}-\mathbf{y})T^e\tilde{\rho}^e(\mathbf{y}) - \frac{1}{2}ig^2 T^eT^e_{ab}\int_{\mathbf{y}}\tilde{\phi}(\mathbf{x}-\mathbf{y})\tilde{\rho}^b(\mathbf{y}) \frac{\delta}{\tilde{\rho}^a(\mathbf{y})} \\
&+\frac{1}{2!} (ig)^2 \int_{\mathbf{y},\mathbf{z}} \tilde{\phi}(\mathbf{x}-\mathbf{y})\tilde{\phi}(\mathbf{x}-\mathbf{z}) T^eT^d\tilde{\rho}^e(\mathbf{y}) \tilde{\rho}^d(\mathbf{z}) +\mathcal{O}(g^3).\\
\end{split}
\end{equation}

\begin{equation}
\begin{split}
&U_R(\mathbf{x}) = \mathrm{exp}\left\{-ig^2\int_{\mathbf{y}} \tilde{\phi}(\mathbf{x}-\mathbf{y}) T^e \mathcal{J}_R^e(\mathbf{y})\right\}\\
=&1- ig\int_{\mathbf{y}} \tilde{\phi}(\mathbf{x}-\mathbf{y})T^e\tilde{\rho}^e(\mathbf{y}) - \frac{1}{2}ig^2 T^eT^e_{ab}\int_{\mathbf{y}}\tilde{\phi}(\mathbf{x}-\mathbf{y})\tilde{\rho}^b(\mathbf{y}) \frac{\delta}{\tilde{\rho}^a(\mathbf{y})} \\
&+\frac{1}{2!} (ig)^2 \int_{\mathbf{y},\mathbf{z}} \tilde{\phi}(\mathbf{x}-\mathbf{y})\tilde{\phi}(\mathbf{x}-\mathbf{z}) T^eT^d\tilde{\rho}^e(\mathbf{y}) \tilde{\rho}^d(\mathbf{z}) +\mathcal{O}(g^3).\\
\end{split}
\end{equation}

At order $\mathcal{O}(g)$ it is obvious that \eq{diamondus} is satisfied, and the first nontrivial check of the relation is at $O(g^2)$. At this order we obtain
\begin{equation}
\begin{split}
& \bar{U}_R(\mathbf{x}) U_L(\mathbf{x}) \bar{U}_L(\mathbf{x}) U_R(\mathbf{x}) =1- ig^2 T^eT^e_{ab} \int_{\mathbf{z}} \tilde{\phi}(\mathbf{x}-\mathbf{z}) \frac{\delta}{\delta \tilde{\rho}^b(\mathbf{x})} \tilde{\rho}^a(\mathbf{z})\\
&-2 ig^2 T^eT^e_{ab} \int_{\mathbf{y},\mathbf{z}} \partial_{\mathbf{y}}\tilde{\phi}(\mathbf{x}-\mathbf{y})\partial_{\mathbf{y}}\tilde{\phi}(\mathbf{y}-\mathbf{z}) \frac{\delta}{\delta \tilde{\rho}^b(\mathbf{y})} \tilde{\rho}^a(\mathbf{z})\\
&+ig\int_{\mathbf{y}} \tilde{\phi}(\mathbf{x}-\mathbf{y})T^a\tilde{\rho}^a(\mathbf{y})gT^b \frac{\delta}{\delta \tilde{\rho}^b(\mathbf{x})} + gT^b \frac{\delta}{\delta \tilde{\rho}^b(\mathbf{x})} (-ig)\int_{\mathbf{y}} \tilde{\phi}(\mathbf{x}-\mathbf{y})T^a\tilde{\rho}^a(\mathbf{y})\\
=&1-2 ig^2 T^eT^e_{ab} \int_{\mathbf{y},\mathbf{z}} \partial_{\mathbf{y}}\tilde{\phi}(\mathbf{x}-\mathbf{y})\partial_{\mathbf{y}}\tilde{\phi}(\mathbf{y}-\mathbf{z}) \frac{\delta}{\delta \tilde{\rho}^b(\mathbf{y})} \tilde{\rho}^a(\mathbf{z})-2ig^2 T^eT^e_{ab} \int_{\mathbf{z}} \tilde{\phi}(\mathbf{x}-\mathbf{z}) \frac{\delta}{\delta \tilde{\rho}^b(\mathbf{x})} \tilde{\rho}^a(\mathbf{z})\ne 1.\\\\
\end{split}
\end{equation}
Thus we have established that at order $\mathcal{O}(g^2)$, the diamond condition is not satisfied by our Wilson line like operators.

We thus conclude that in spite of  certain similarities, the self dual RFT Hamiltonian \eq{ha}  is not the same as the effective action of \cite{diamond}.  The status of this comparison is further discussed in the next section.

\section{Discussion}
In this paper we have revisited the problem of constructing a self dual Reggeon Field Theory Hamiltonian $H_{RFT}$. We have followed the EFT strategy by imposing the relevant symmetries and also required that $H_{RFT}$ reduces to $H_{JIMWLK}$ (or $H_{KLWMIJ}$) in the dense-dilute limit. 

As a result we have found a family of Hamiltonians that satisfy these requirements. These Hamiltonians are constructed from Wilson line - like operators in different representations of the $SU(N)$ group. 
  We note that any of these Hamiltonians in addition to reproducing the dense dilute limit, also generates correct Pomeron loops. The simplest way to see this is to perform the coupling constant expansion using the BFKL counting introduced in Section 5. This is equivalent to simultaneous expansion  in powers of $\rho$ and $\delta/\delta\rho$. At order $\alpha_s$ the Hamiltonian reduced to $H_{BFKL}$, while at order $\alpha_s^2$ it contains both splitting and merging  vertices ($\rho^2 (\delta/\delta\rho )^4$ and $\rho^4 (\delta/\delta\rho )^2$) with correct coefficients. As discussed in \cite{pomloops,ddd} these vertices are responsible both for a certain set of reggeization corrections, and for the QCD Pomeron loops.  

We have analyzed the continuous symmetries of $H_{RFT}$. This is an interesting question since both $H_{JIMWLK}$ and $H_{KLWMIJ}$ possess an $SU_L(N)\times SU_R(N)$ symmetry group, but the generators of these transformations are not the same in the two dense-dilute cases. For $H_{RFT}$ we are able to show nonperturbatively the existence of one $SU_V(N)$ symmetry, which is the diagonal subgroup of the symmetry group in both JIMWLK and KLWMIJ limits. We established the fact that the two diagonal subgroups are identical explicitly using the algebra of the generators in the RFT Hilbert space.
We have also shown that $H_{RFT}$ is invariant under the left and right rotations at least to $O(g^6)$ in perturbative expansion. This is a strong indication that the continuous symmetry group is at least  $SU(N)\times SU(N)\times SU(N)$.

 One member of the family of the Hamiltonians we found is very similar to the "diamond action"\cite{diamond,Balitsky05}. Our Hamiltonian RFT framework is different from the effective action approach of \cite{diamond,Balitsky05} which somewhat hampers direct comparison. Nevertheless if we juxtapose our $H_{RFT}$ defined in terms of adjoint Wilson lines directly with the effective action of  \cite{diamond,Balitsky05}, the two look identical. There is however one significant difference between our result and that of \cite{diamond}. Namely the action in \cite{diamond} is written in terms of four Wilson loops that satisfy the diamond condition, \eq{diamondc}. This condition played a very important role in \cite{diamond}. In fact the effective action derived in \cite{diamond} directly from QCD is equivalent to the "KLWMIJ+" Hamiltonian suggested in \cite{KLremark,KLremark2}, whereby KLWMIJ Hamiltonian is generalized by including nonlinear corrections in the solution for classical field.  This Hamiltonian is not explicitly self dual, and only with the help of the diamond condition it was  recast in \cite{diamond} in the form which looks self dual, at least superficially. 
However whether the "diamond action" is in fact self dual or not remained an open question. To check the self duality one has to  verify that the duality transformation is canonical, or in the quantum sense a linear transformation on the RFT Hilbert space. This was not possible to do with the tools of \cite{diamond}, as no operator realization of the algebra of Wilson lines was explicitly presented. In the present paper we operate within the RFT Hilbert space with well defined operator algebra; and therefore we have explicit realization of the duality transformation in the Hilbert space.  We find within this consistent framework that the diamond action (RFT Hamiltonian) is self dual, but the diamond condition between the Wilson lines is not satisfied. The condition is violated starting with order $O(g^2)$ in perturbative expansion. In this sense our paper is closer to \cite{Balitsky05}, where the diamond action is derived as a self dual form of the action in the dense-dilute limit without assuming the diamond constraint between the Wilson lines. In \cite{Balitsky05} the constraint  was shown to hold in the first order in perturbation theory, which is consistent with our conclusion here, but was not checked at higher orders.

 Our "bottom up" approach does not allow us to decide which one of the candidate hamiltonians we have found is the right one, and in fact whether any one of them is the correct QCD RFT Hamiltonian. 
 Even though we have used the EFT methodology to determine possible terms in $H_{RFT}$, we are at a disadvantage here compared to standard applications of EFT in quantum field theory. The generic situation is that one is searching for local operators that can be incorporated into the EFT Lagrangian (or Hamiltonian) in the situation where there is only a finite number of possible operators of a given dimension. The higher the dimension of the operator the stronger the suppression of  its contribution to low energy observables. Thus EFT organizes the possible operators according to their importance in the interesting kinematics. In our case the situation appears to be different. Although RFT is the effective theory of QCD at high energy, all the operators we have found may contribute at  leading order in $E^{-1}$. We do not see any obvious parameter which would order the possible contributions. 
 The similarity with the diamond action may suggest that one should work with the Wilson lines in the adjoint representation. However as is clear from the derivation in \cite{Balitsky05} the diamond action is not the full story, but is only a leading term in an expansion away from the abelian limit. Thus it is possible that the other candidate terms we have found also play a role in the full $RFT$ Hamiltonian.  
 
 It would be interesting to find a criterion which could discriminate between the possible terms. One possibility is to compare $H_{RFT}$ with NLO JIMWLK. Although we have no reason to expect that $H_{RFT}$ contains all, or even most NLO terms, it does contain some such terms. Comparing those to NLO JIMWLK could be instructive and possibly discriminatory. 

 Another interesting question is the unitarity of $H_{RFT}$. 
 As we have mentioned in the introduction, our main motivation to search for the self dual $H_{RFT}$ was the unitarity violation in $H_{JIMWLK}$. 
   The question of unitary really has two parts: the t-channel unitarity and the s-channel unitarity. 
  
  Although we have not studied this in detail here, it is broadly believed that the t-channel unitarity, which has been a cornerstone 
 of Gribov's RFT is ensured by the self-duality of $H_{RFT}$.  This connection is rooted in boost invariance of the scattering amplitudes. On one hand Lorentz invariance requires  self-duality of RFT \cite{KLduality}, and  at the same time,   boost invariance has been argued  to be equivalent to the  t-channel unitarity, see \cite{Kancheli} for  latest discussion. On the technical level we note that  the coupling constant  expansion of  $H_{RFT}$ (in the BFKL counting discussed above) in the large $N_c$ limit
 generates the Gribov Pomeron calculus. Scattering amplitudes are  then represented in terms of the exchanges of the BFKL Pomerons and their interactions via the "merging" and "splitting" three Pomeron vertexes. Such a theory is known to satisfy the $t$-channel unitarity, and we are therefore confident that our $H_{RFT}$ indeed is t-channel unitary.

 As for the s-channel unitarity, the situation here is more complex. We have formulated the conditions for s-channel unitarity in \cite{KLLL}. 
 Given $H_{RFT}$ one can in principle follow the procedure explained in \cite{KLLL}  to determine whether its action corresponds to unitarity evolution of QCD states in energy. 
  
 This entails taking a generic QCD projectile state  
$$
|\Psi_i\rangle_P= C^{in}_{b_1,b_2...b_m}|\mathbf{y}_1,b_1;...;\mathbf{y}_m,b_m\rangle
$$
and evolving it to infinitesimally higher energy.  The result of the evolution in general can be represented in the form:
$$
\,|\Psi\rangle \,\rightarrow\, \sum_{n; \mathbf{x}_i;a_i}C_{a_1,a_2...a_n}|\mathbf{x}_1,a_1;...;\mathbf{x}_n,a_n\rangle
$$ 
The energy evolution of the scattering amplitude of this evolved state on a fixed target is given by the action of $H_{RFT}$ as in eq.(\ref{sev}).
Next, one has to construct a probability function  $F$ defined in (\ref{fm}) and 
verify the unitarity condition (\ref{prob}). The unitarity should hold for any initial state $|\Psi_i\rangle_P$.  
 
 In principle one should be able to pursue this calculation, since the algebra of RFT is explicitly known, and therefore the action of $H_{RFT}$ on an unevolved amplitude is completely defined.
  Unfortunately analyzing the unitarity conditions beyond the JIMWLK limit is technically a complicated problem,  due to complicated algebra of the Wilson lines, which at this point we are not able to solve. We believe it is a very important question and are planning to address it in future work.

\section{Acknowledgements}
   We thank our colleagues at Tel Aviv university and UTFSM for
 encouraging discussions.     AK and Ming Li were supported by the NSF Nuclear Theory grants 1614640 and 1913890.
  EL was supported  by 
 ANID PIA/APOYO AFB180002 (Chile) and  Fondecyt (Chile) grant  
\# 1180118.  
ML was supported by the Israeli Science Foundation (ISF) grant \#1635/16.
ML and AK were also supported by the Binational Science Foundation grants \#2015626, \#2018722, and the Horizon 2020 RISE
 "Heavy ion collisions: collectivity and precision in saturation physics"  under grant agreement No. 824093.
This work has been performed
in the framework of COST Action CA15213 ``Theory of hot matter and relativistic heavy-ion collisions" (THOR).

\appendix
\section{Checking $SU_L(N)$.}
In this Appendix we calculate perturbatively the commutator of $Q^a_L$ with the Hamiltonian. 

\subsection{$[Q^a_L,\bar V_L(\mathbf{x})]$.}
We start by trying to verify the conjectured commutation relation:
\begin{equation}\label{commu}
\left[Q_L^a, \bar{V}_L^{\alpha\beta}(\mathbf{x})\right] = -\left(t^a\bar{V}_L(\mathbf{x})\right)^{\alpha\beta} (???)
\end{equation}

We calculate the commutator perturbatively using the BFKL counting.  We express 
\begin{equation}
\mathcal{J}_L^a(z) = \frac{1}{g} B^a_{(-1)}(\mathbf{z}) + B^a_{(0)}(\mathbf{z}) + g B^a_{(1)}(\mathbf{z}) + g^3 B^a_{(3)}(\mathbf{z}) + \ldots 
\end{equation}
with 
\begin{equation}
\begin{split}
&B^a_{(-1)} = \tilde{\rho}^a(\mathbf{z})\, ,\\
&B^a_{(0)}  = -\frac{1}{2} \tilde{\rho}^b(\mathbf{z}) T^e_{ba} \frac{\delta}{\delta \tilde{\rho}^e(\mathbf{z})}\, ,\\
&B^a_{(1)} = \frac{1}{12} \tilde{\rho}^b_{\mathbf{z}} (T^{e_1}T^{e_2})_{ba} \frac{\delta}{\delta \tilde{\rho}^{e_1}_{\mathbf{z}}} \frac{\delta}{\delta \tilde{\rho}^{e_2}_{\mathbf{z}}} ,\\
&B^a_{(3)} = -\frac{1}{720} \tilde{\rho}^b_{\mathbf{z}} (T^{e_1}T^{e_2}T^{e_3}T^{e_4})_{ba} \frac{\delta}{\delta \tilde{\rho}^{e_1}_{\mathbf{z}}} \frac{\delta}{\delta \tilde{\rho}^{e_2}_{\mathbf{z}}}\frac{\delta}{\delta \tilde{\rho}^{e_3}_{\mathbf{z}}} \frac{\delta}{\delta \tilde{\rho}^{e_4}_{\mathbf{z}}}, \\
\end{split}
\end{equation}
To expand $\bar V_L$ we need 
\begin{equation}
\mathcal{I}_L^e(\mathbf{y}) = \frac{1}{g} E_{(-1)}(\mathbf{y}) + E_{(0)}(\mathbf{y})  + gE_{(1)}(\mathbf{y}) + g^3 E_{(3)}(\mathbf{y}) + \ldots 
\end{equation}
with
\begin{equation}
\begin{split}
&E^a_{(-1)} = -i \partial^2_{\mathbf{y}} \frac{\delta}{\delta \tilde{\rho}^a_{\mathbf{y}}}\\
&E^a_{(0)} =-\frac{1}{2} \partial^2_{\mathbf{y}} \frac{\delta}{\delta \tilde{\rho}^b_{\mathbf{y}}} T^e_{ba} \frac{1}{\partial^2} \tilde{\rho}^e_{\mathbf{y}} \\
&E^a_{(1)} = \frac{i}{12}  \partial^2_{\mathbf{y}} \frac{\delta}{\delta \tilde{\rho}^b_{\mathbf{y}}} (T^{e_1}T^{e_2})_{ba} \frac{1}{\partial^2}\tilde{\rho}_{\mathbf{y}}^{e_1} \frac{1}{\partial^2} \tilde{\rho}^{e_2}_{\mathbf{y}} \\
&E^a_{(3)} = \frac{i}{720} \partial^2_{\mathbf{y}} \frac{\delta}{\delta \tilde{\rho}^b_{\mathbf{y}}} (T^{e_1}T^{e_2}T^{e_3}T^{e_4})_{ba} \frac{1}{\partial^2}\tilde{\rho}_{\mathbf{y}}^{e_1} \frac{1}{\partial^2} \tilde{\rho}_{\mathbf{y}}^{e_2} \frac{1}{\partial^2} \tilde{\rho}_{\mathbf{y}}^{e_3} \frac{1}{\partial^2} \tilde{\rho}_{\mathbf{y}}^{e_4} .
\end{split}
\end{equation}
Then  $\bar{V}_L$ is expanded as
\begin{equation}
\begin{split}
\bar{V}_L = &e^{ gA_{(1)} + g^2 A_{(2)} + g^3 A_{(3)} +g^5 A_{(5)}+\ldots}\\
=&1 + gA_{(1)} + g^2 \left(A_{(2)} + \frac{1}{2} A_{(1)}A_{(1)} \right)+ g^3\left(A_{(3)} + \frac{1}{2}( A_{(1)}A_{(2)} + A_{(2)}A_{(1)})+\frac{1}{3!} (A_{(1)})^3  \right)\\
&+g^4\Big(\frac{1}{2} (A_{(2)})^2 + \frac{1}{2}(A_{(1)}A_{(3)} + A_{(3)}A_{(1)}) + \frac{1}{3!}(A_{(1)}A_{(1)}A_{(2)} + A_{(1)}A_{(2)} A_{(1)} + A_{(2)}A_{(1)}A_{(1)})\\
&\qquad \qquad +\frac{1}{4!} (A_{(1)})^4\Big) + \ldots 
\end{split}
\end{equation}
with 
\begin{equation}
A_{(1)} = i\int_{\mathbf{y}} \tilde{\phi}(\mathbf{x}-\mathbf{y}) t^e \left(-i \partial^2_{\mathbf{y}} \frac{\delta}{\delta \tilde{\rho}^e_{\mathbf{y}}}\right) = t^e \frac{\delta}{\delta \tilde{\rho}^e_{\mathbf{x}}}
\end{equation}
\begin{equation}
\begin{split}
A_{(2)} =& i\int_{\mathbf{y}} \tilde{\phi}(\mathbf{x}-\mathbf{y}) t^d \left(-\frac{1}{2} \partial^2_{\mathbf{y}} \frac{\delta}{\delta \tilde{\rho}^b_{\mathbf{y}}} T^e_{bd} \frac{1}{\partial^2} \tilde{\rho}^e(\mathbf{y}) \right)\\
=&\frac{i}{2} (t^et^b-t^bt^e) \Big( \frac{\delta}{\delta\tilde{\rho}^b_{\mathbf{x}}} \int_{\mathbf{w}} \tilde{\phi}(\mathbf{x} -\mathbf{w}) \tilde{\rho}^e_{\mathbf{w}} + \int_{\mathbf{y}}\tilde{\phi}(\mathbf{x}-\mathbf{y}) \frac{\delta}{\delta\tilde{\rho}^b_{\mathbf{y}}}\tilde{\rho}^e_{\mathbf{y}}\\
&+2\int_{\mathbf{y}} \partial_{\mathbf{y}}\tilde{\phi}(\mathbf{x}-\mathbf{y}) \frac{\delta}{\delta \tilde{\rho}^b_{\mathbf{y}}} \int_{\mathbf{w}} \partial_{\mathbf{y}}\tilde{\phi}(\mathbf{y} -\mathbf{w}) \tilde{\rho}^e_{\mathbf{w}}\Big)\\
\end{split}
\end{equation}
\begin{equation}
\begin{split}
A_{(3)} =  &i\int_{\mathbf{y}} \tilde{\phi}(\mathbf{x}-\mathbf{y}) t^d\left(\frac{i}{12}  \partial^2_{\mathbf{y}} \frac{\delta}{\delta \tilde{\rho}^b} (T^{e_1}T^{e_1})_{bd} \frac{1}{\partial^2}\tilde{\rho}^{e_1} \frac{1}{\partial^2} \tilde{\rho}^{e_2} \right)\\
=&-\frac{1}{12} (t^{e_1}t^{e_2}t^b - 2t^{e_1}t^bt^{e_2} + t^bt^{e_1}t^{e_2}) \int_{\mathbf{y}}\tilde{\phi}(\mathbf{x}-\mathbf{y})\partial^2_{\mathbf{y}} \frac{\delta}{\delta \tilde{\rho}^b_{\mathbf{y}}}\int_{\mathbf{z_1}} \tilde{\phi}(\mathbf{y}-\mathbf{z}_1) \tilde{\rho}^{e_1}_{\mathbf{z}_1} \int_{\mathbf{z_2}} \tilde{\phi}(\mathbf{y}-\mathbf{z}_2) \tilde{\rho}^{e_2}_{\mathbf{z}_2} .\\
\end{split}
\end{equation}

In terms of coupling constant $g$, we check the commutator \eq{commu} order by order.

\begin{itemize}
\item 
$O(g^0)$ is  satisfied.
\begin{equation}
\left[\int_{\mathbf{z}} B^a_{(-1)}(\mathbf{z}), A_{(1)}(\mathbf{x}) \right]  = -t^a. 
\end{equation}

\item
 $O(g)$, the relation to be checked is 
\begin{equation}\label{eq:check_g1}
\left[\int_{\mathbf{z}} B^a_{(-1)}(\mathbf{z}), A_{(2)}+\frac{1}{2} (A_{(1)} )^2\right]  + \left[\int_{\mathbf{z}} B^a_{(0)}(\mathbf{z}), A_{(1)}(\mathbf{x}) \right]  =-t^a A_{(1)}.
\end{equation}

First note that each individual term is 
\begin{equation}
\begin{split}
\left[\int_{\mathbf{z}} B^a_{(0)}(\mathbf{z}), A_{(1)}(\mathbf{x}) \right]  &=   \frac{1}{2} t^b T^e_{ba}  \frac{\delta}{\delta \tilde{\rho}^e_{\mathbf{x}}} = -\frac{1}{2} (t^at^e -t^et^a) \frac{\delta}{\delta \tilde{\rho}^e_{\mathbf{x}}} \\
& = -\frac{1}{2} \left(t^aA_{(1)}-A_{(1)}t^a\right)\\
& = -\frac{1}{2} \left[t^a, A_{(1)}\right].
\end{split}
\end{equation}
\begin{equation}
\begin{split}
\left[\int_{\mathbf{z}} B^a_{(-1)}(\mathbf{z}), \frac{1}{2} (A_{(1)} )^2\right]  =& -\frac{1}{2}  (t^{e} t^a + t^a t^{e}) \frac{\delta}{\delta \tilde{\rho}^e_{\mathbf{x}}} \\
=&-\frac{1}{2} \left(t^aA_{(1)} + A_{(1)}t^a\right).\\
\end{split}
\end{equation}
We have one additional term but it vanishes.
\begin{equation}\label{eq:vanishing}
\left[\int_{\mathbf{z}} B^a_{(-1)}(\mathbf{z}), A_{(2)}\right]   = -\int_{\mathbf{z}}(-\frac{i}{2}) t^d T^e_{bd} \int_{\mathbf{y}} \tilde{\phi}(\mathbf{x}-\mathbf{y})[ \partial^2_{\mathbf{y}}\delta(\mathbf{y}-\mathbf{z}) \delta^{ba} ]\int_{\mathbf{w}} \tilde{\phi}(\mathbf{y} -\mathbf{w}) \tilde{\rho}^e_{\mathbf{w}} =0.
\end{equation}
This vanishes due to 
$\partial_{\mathbf{y}}^2 \int_{\mathbf{z}} \delta(\mathbf{y}-\mathbf{z}) =0$.

\item
 $O(g^2)$, the relation to be checked is 
\begin{equation}\label{eq:g2_order_relation}
\begin{split}
&\int_{\mathbf{z}} \left[B_{(-1)}^a(\mathbf{z}), A_{(3)} + \frac{1}{2}( A_{(1)}A_{(2)} + A_{(2)}A_{(1)})+\frac{1}{3!} (A_{(1)})^3  \right] \\
&+\int_{\mathbf{z}} \left[B_{(0)}(\mathbf{z}), A_{(2)} + \frac{1}{2} A_{(1)}A_{(1)} \right]+\int_{\mathbf{z}} \left[B_{(1)}(\mathbf{z}), A_{(1)}\right]\\
&=-t^a\left(A_{(2)} +\frac{1}{2} A_{(1)} A_{(1)}\right).\\
\end{split}
\end{equation}
First note that
\begin{equation}
\begin{split}
&\int_{\mathbf{z}} \left[B_{(-1)}^a(\mathbf{z}), A_{(3)}\right] =0\,, \\
&\int_{\mathbf{z}} \left[B_{(-1)}^a(\mathbf{z}), A_{(2)}\right] =0\, .\\
\end{split}
\end{equation}
for the same reason as \eq{eq:vanishing}. This is obviously a general property. 
Now we evaluate each term. 
\begin{equation}\label{eq:four}
\int_{\mathbf{z}} \left[B_{(-1)}^a(\mathbf{z}),  \frac{1}{2}( A_{(1)}A_{(2)} + A_{(2)}A_{(1)})\right] = - \frac{1}{2} (t^a A_{(2)} + A_{(2)} t^a)
\end{equation}
\begin{equation}\label{eq:one}
\begin{split}
\int_{\mathbf{z}} \left[B_{(-1)}^a(\mathbf{z}),\frac{1}{3!} (A_{(1)})^3  \right]  = &-\frac{1}{6} \left(t^at^{e_1}t^{e_2} + t^{e_1}t^at^{e_2}+t^{e_1}t^{e_2}t^a\right)\frac{\delta}{\delta \tilde{\rho}^{e_1}_{\mathbf{x}}}\frac{\delta}{\delta \tilde{\rho}^{e_2}_{\mathbf{x}}}\\
=&-\frac{1}{6}(t^aA_{(1)}A_{(1)} + A_{(1)}t^aA_{(1)} + A_{(1)}A_{(1)}t^a)\\
\end{split}
\end{equation}
\begin{equation}\label{eq:two}
\begin{split}
\int_{\mathbf{z}} \left[B_{(1)}(\mathbf{z}), A_{(1)}\right] = &-\frac{1}{12} t^e( T^{e_1}T^{e_2})_{ea} \frac{\delta}{\delta \tilde{\rho}^{e_1}_{\mathbf{x}}}\frac{\delta}{\delta \tilde{\rho}^{e_2}_{\mathbf{x}}}\\
=&-\frac{1}{12} (t^at^{e_1}t^{e_2} - 2t^{e_1}t^at^{e_2} + t^{e_1}t^{e_2}t^a) \frac{\delta}{\delta \tilde{\rho}^{e_1}_{\mathbf{x}}}\frac{\delta}{\delta \tilde{\rho}^{e_2}_{\mathbf{x}}}\\
=&-\frac{1}{12} (t^aA_{(1)}A_{(1)} -2A_{(1)}t^aA_{(1)} +A_{(1)}A_{(1)}t^a)\\
=&-\frac{1}{12} \left[\left[t^a, A_{(1)}\right], A_{(1)}\right]
\end{split}
\end{equation}
\begin{equation}\label{eq:three}
\begin{split}
\int_{\mathbf{z}} \left[B_{(0)}(\mathbf{z}),  \frac{1}{2} A_{(1)}A_{(1)} \right] =& -\frac{1}{4} (t^{a}t^{e_1}t^{e_2} - t^{e_1}t^{e_2}t^a)  \frac{\delta}{\delta \tilde{\rho}^{e_1}_{\mathbf{x}}}\frac{\delta}{\delta \tilde{\rho}^{e_2}_{\mathbf{x}}}\\
=&-\frac{1}{4}(t^aA_{(1)}A_{(1)}-A_{(1)}A_{(1)}t^a).
\end{split}
\end{equation}
Adding Eqs. \eqref{eq:one}, \eqref{eq:two}, \eqref{eq:three}, one obtains 
\begin{equation}
-\frac{1}{2} (t^at^{e_1}t^{e_2})  \frac{\delta}{\delta \tilde{\rho}^{e_1}_{\mathbf{x}}}\frac{\delta}{\delta \tilde{\rho}^{e_2}_{\mathbf{x}}} =  -t^a \left(\frac{1}{2} A_{(1)}A_{(1)}\right)
\end{equation}
which is part of the right hand side of the relation \eq{eq:g2_order_relation}. To continue
\begin{equation}\label{eq:five}
\begin{split}
\int_{\mathbf{z}} \left[B_{(0)}(\mathbf{z}), A_{(2)} \right] =& \int_{\mathbf{z}}\left[-\frac{1}{2} \tilde{\rho}^p_{\mathbf{z}} T^q_{pa} \frac{\delta}{\delta \tilde{\rho}^q_{\mathbf{z}}},  -\frac{i}{2} t^d T^e_{bd} \int_{\mathbf{y}} \tilde{\phi}(\mathbf{x}-\mathbf{y}) \partial^2_{\mathbf{y}} \frac{\delta}{\delta \tilde{\rho}^b_{\mathbf{y}}} \int_{\mathbf{w}} \tilde{\phi}(\mathbf{y} -\mathbf{w}) \tilde{\rho}^e_{\mathbf{w}}\right] \\
=&\left(-\frac{i}{4} T^{m}_{pa} T^n_{pd}+ \frac{i}{4} T^e_{na} T^e_{md} \right) t^d \int_{\mathbf{y}} \tilde{\phi}(\mathbf{x}-\mathbf{y}) \partial^2_{\mathbf{y}} \frac{\delta}{\delta \tilde{\rho}^m_{\mathbf{y}}}  \int_{\mathbf{w}} \tilde{\phi}(\mathbf{y} -\mathbf{w}) \tilde{\rho}^n_{\mathbf{w}} \\
=&\frac{1}{2} (A_{(2)}t^a -t^aA_{(2)}). \\
\end{split}
\end{equation}
So Eqs. \eqref{eq:four}, \eqref{eq:five} adds up gives $-t^a A_{(2)}$, which is exactly the last piece on the right hand side of Eq. \eqref{eq:g2_order_relation}.  To second order in $g$ \eq{commu} holds.

\item 
$O(g^3)$. 

The relation to be proved is 
\begin{equation}\label{eq:g3_order_relation}
\begin{split}
&\int_{\mathbf{z}} \Big[ B_{(-1)}^a(\mathbf{z}), \frac{1}{2} ( (A_{(2)})^2 + A_{(1)}A_{(3)} + A_{(3)}A_{(1)}) \\
&\qquad + \frac{1}{3!}( A_{(1)}A_{(1)}A_{(2)} + A_{(1)} A_{(2)}A_{(1)}+A_{(2)}A_{(1)}A_{(1)})+ \frac{1}{4!} (A_{(1)})^4\Big]\\
+&\int_{\mathbf{z}} \Big[ B_{(0)}^a(\mathbf{z}),  A_{(3)} + \frac{1}{2}(A_{(1)}A_{(2)} + A_{(2)}A_{(1)}) + \frac{1}{3!}(A_{(1)})^3\Big]\\
+&\int_{\mathbf{z}} \Big[ B_{(1)}^a(\mathbf{z}), A_{(2)} + \frac{1}{2}A_{(1)}A_{(1)}\Big]\\
=& -t^a\left(A_{(3)} + \frac{1}{2}(A_{(1)}A_{(2)} + A_{(2)}A_{(1)}) + \frac{1}{3!}(A_{(1)})^3\right).
\end{split}
\end{equation}
We calculate each commutator separately.  The first one is easy to compute as we know that 
\begin{equation}
\begin{split}
&\int_{\mathbf{z}} \Big[ B_{(-1)}^a(\mathbf{z}), A_{(1)}\Big] = -t^a, \\ 
&\int_{\mathbf{z}} \Big[ B_{(-1)}^a(\mathbf{z}), A_{(n)}\Big] =0, \quad \mathrm{for}\,\,\, n\geq 2\\
\end{split}
\end{equation}
Using this relation, one obtains
\begin{equation}
\begin{split}
&\int_{\mathbf{z}} \Big[ B_{(-1)}^a(\mathbf{z}), \frac{1}{2} ( (A_{(2)})^2 + A_{(1)}A_{(3)} + A_{(3)}A_{(1)}) \\
&\qquad + \frac{1}{3!}( A_{(1)}A_{(1)}A_{(2)} + A_{(1)} A_{(2)}A_{(1)}+A_{(2)}A_{(1)}A_{(1)})+ \frac{1}{4!} (A_{(1)})^4\Big]\\
=&-\frac{1}{2} \Big(t^aA_{(3)} + A_{(3)}t^a\Big) -\frac{1}{6}\Big(t^aA_{(1)}A_{(2)} + A_{(1)}t^aA_{(2)} + t^aA_{(2)}A_{(1)} + A_{(1)}A_{(2)}t^a \\
&\qquad + A_{(2)}t^aA_{(1)} + A_{(2)}A_{(1)}t^a\Big)-\frac{1}{24}\Big(t^aA_{(1)}A_{(1)}A_{(1)} + A_{(1)}t^aA_{(1)}A_{(1)} \\
&\qquad + A_{(1)}A_{(1)}t^aA_{(1)} + A_{(1)}A_{(1)}A_{(1)}t^a\Big). 
\end{split}
\end{equation}
We also notice that
\begin{equation}
\int_{\mathbf{z}} \Big[ B_{(0)}^a(\mathbf{z}),  A_{(n)} \Big] = \frac{1}{2} (A_{(n)}t^a - t^aA_{(n)}).
\end{equation}
for $n=1,2, 3$. It is possible that this relation holds for all the relevant $n$. Using this relation, we calculate
\begin{equation}
\begin{split}
&\int_{\mathbf{z}} \Big[ B_{(0)}^a(\mathbf{z}),  A_{(3)} + \frac{1}{2}(A_{(1)}A_{(2)} + A_{(2)}A_{(1)}) + \frac{1}{3!}(A_{(1)})^3\Big]\\
=&\frac{1}{2}(A_{(3)}t^a - t^aA_{(3)}) + \frac{1}{4} (A_{(1)}t^a-t^aA_{(1)})A_{(2)} + \frac{1}{4} A_{(1)}(A_{(2)}t^a-t^aA_{(2)}) \\
&+ \frac{1}{4} (A_{(2)}t^a-t^aA_{(2)})A_{(1)} + \frac{1}{4} A_{(2)}(A_{(1)}t^a -t^aA_{(1)}) + \frac{1}{12} (A_{(1)}t^a-t^aA_{(1)})A_{(1)}A_{(1)}\\
&+\frac{1}{12} A_{(1)}(A_{(1)}t^a-t^aA_{(1)}) A_{(1)} +\frac{1}{12}A_{(1)}A_{(1)}(A_{(1)}t^a-t^aA_{(1)})\\
=&\frac{1}{2}(A_{(3)}t^a - t^aA_{(3)})  +\frac{1}{4}\Big(A_{(1)}A_{(2)}t^a+A_{(2)}A_{(1)}t^a - t^aA_{(1)}A_{(2)} -t^aA_{(2)}A_{(1)}\Big) \\
&+\frac{1}{12}\Big(A_{(1)}A_{(1)}A_{(1)} t^a -t^a A_{(1)}A_{(1)}A_{(1)}\Big).\\
\end{split}
\end{equation}
From Eq. \eqref{eq:two}, we know that
\begin{equation}
\int_{\mathbf{z}} \left[B_{(1)}(\mathbf{z}), A_{(1)}\right] =  -\frac{1}{12}(t^aA_{(1)}A_{(1)} -2A_{(1)}t^aA_{(1)} + A_{(1)}A_{(1)}t^a). 
\end{equation}
Usig this relation, one can compute
\begin{equation}
\begin{split}
&\int_{\mathbf{z}} \Big[ B_{(1)}^a(\mathbf{z}),  \frac{1}{2}A_{(1)}A_{(1)}\Big]\\
=&-\frac{1}{24} \Big(t^aA_{(1)}A_{(1)}A_{(1)} -2A_{(1)}t^aA_{(1)} A_{(1)}+ A_{(1)}A_{(1)}t^aA_{(1)} \\
&\qquad +A_{(1)}t^aA_{(1)}A_{(1)} -2A_{(1)}A_{(1)}t^aA_{(1)} +A_{(1)} A_{(1)}A_{(1)}t^a\Big).\\
=&-\frac{1}{24} \Big(t^aA_{(1)}A_{(1)}A_{(1)} -A_{(1)}t^aA_{(1)} A_{(1)} -A_{(1)}A_{(1)}t^aA_{(1)} +A_{(1)} A_{(1)}A_{(1)}t^a\Big).\\
\end{split}
\end{equation}
The last piece we need to calculate is 
\begin{equation}
\begin{split}
&\int_{\mathbf{z}} \Big[ B_{(1)}^a(\mathbf{z}), A_{(2)}(\mathbf{x}) \Big]\\
=&\int_{\mathbf{z}}\Bigg[\frac{1}{12} \tilde{\rho}^c_{\mathbf{z}} (T^{e_1}T^{e_2})_{ca} \frac{\delta}{\delta \tilde{\rho}^{e_1}_{\mathbf{z}}} \frac{\delta}{\delta \tilde{\rho}^{e_2}_{\mathbf{z}}}\, , \,\frac{i}{2} (t^et^b-t^bt^e) \int_{\mathbf{y}} \tilde{\phi}(\mathbf{x}-\mathbf{y}) \partial^2_{\mathbf{y}} \frac{\delta}{\delta \tilde{\rho}^b_{\mathbf{y}}} \int_{\mathbf{w}} \tilde{\phi}(\mathbf{y} -\mathbf{w}) \tilde{\rho}^e_{\mathbf{w}}\Bigg]\\
=&\frac{1}{12}(T^{e_1}T^{e_2})_{ba}\frac{i}{2}T^p_{be}t^p\left(-\int_{\mathbf{z}}\frac{\delta}{\delta \tilde{\rho}^{e_1}_{\mathbf{z}}} \frac{\delta}{\delta \tilde{\rho}^{e_2}_{\mathbf{z}}}\int_{\mathbf{y}}\tilde{\phi}(\mathbf{x}-\mathbf{y}) \partial_{\mathbf{y}}^2 \delta(\mathbf{y}-\mathbf{z}) \int_{\mathbf{w}} \tilde{\phi}(\mathbf{y} -\mathbf{w}) \tilde{\rho}^e_{\mathbf{w}}\right)\\
&+\frac{1}{12}(T^{e}T^{d} + T^{d}T^e)_{ha}\frac{i}{2}T^p_{be}t^p\left(\int_{\mathbf{z}} \tilde{\rho}_{\mathbf{z}}^h \frac{\delta}{\delta \tilde{\rho}^d_{\mathbf{z}}}\int_{\mathbf{y}} \tilde{\phi}(\mathbf{x}-\mathbf{y}) \partial^2_{\mathbf{y}} \frac{\delta}{\delta \tilde{\rho}^b_{\mathbf{y}}} \tilde{\phi}(\mathbf{y} -\mathbf{z})\right)\\
=&\frac{1}{12}(T^{b}T^{d})_{qa}\frac{i}{2}T^p_{qh}t^p\left(-\int_{\mathbf{z}}\frac{\delta}{\delta \tilde{\rho}^{b}_{\mathbf{z}}} \frac{\delta}{\delta \tilde{\rho}^{d}_{\mathbf{z}}}\int_{\mathbf{y}}\tilde{\phi}(\mathbf{x}-\mathbf{y}) \partial_{\mathbf{y}}^2 \delta(\mathbf{y}-\mathbf{z}) \int_{\mathbf{w}} \tilde{\phi}(\mathbf{y} -\mathbf{w}) \tilde{\rho}^h_{\mathbf{w}}\right)\\
&+\frac{1}{12}(T^{e}T^{d} + T^{d}T^e)_{ha}\frac{i}{2}T^p_{be}t^p\left(\int_{\mathbf{z}} \tilde{\rho}_{\mathbf{z}}^h \frac{\delta}{\delta \tilde{\rho}^d_{\mathbf{z}}}\int_{\mathbf{y}} \tilde{\phi}(\mathbf{x}-\mathbf{y}) \partial^2_{\mathbf{y}} \frac{\delta}{\delta \tilde{\rho}^b_{\mathbf{y}}} \tilde{\phi}(\mathbf{y} -\mathbf{z})\right)\\
=&\frac{i}{24} \left( - (T^pT^bT^d)_{ha} t^p \right)\Big(-\frac{\delta}{\delta \tilde{\rho}^{b}_{\mathbf{x}}} \frac{\delta}{\delta \tilde{\rho}^{d}_{\mathbf{x}}}  \int_{\mathbf{w}} \tilde{\phi}(\mathbf{x} -\mathbf{w}) \tilde{\rho}^h_{\mathbf{w}} -\int_{\mathbf{z}}\tilde{\phi}(\mathbf{x}-\mathbf{z})\frac{\delta}{\delta \tilde{\rho}^{b}_{\mathbf{z}}} \frac{\delta}{\delta \tilde{\rho}^{d}_{\mathbf{z}}}\tilde{\rho}^h_{\mathbf{z}}\\
&-2\int_{\mathbf{z}}\frac{\delta}{\delta \tilde{\rho}^{b}_{\mathbf{z}}} \frac{\delta}{\delta \tilde{\rho}^{d}_{\mathbf{z}}}\partial_{\mathbf{z}}\tilde{\phi}(\mathbf{x}-\mathbf{z})  \int_{\mathbf{w}} \partial_{\mathbf{z}}\tilde{\phi}(\mathbf{z} -\mathbf{w}) \tilde{\rho}^h_{\mathbf{w}}\Big)\\
&+\frac{i}{24}\left( (T^pT^hT^a)_{bd}t^p + (T^pT^aT^h)_{bd}t^p\right)\Bigg(\frac{\delta}{\delta \tilde{\rho}^b_{\mathbf{x}}}\int_{\mathbf{z}} \tilde{\phi}(\mathbf{x}-\mathbf{z})\tilde{\rho}_{\mathbf{z}}^h \frac{\delta}{\delta \tilde{\rho}^d_{\mathbf{z}}}  \\
&+ \int_{\mathbf{z}}\tilde{\phi}(\mathbf{x}-\mathbf{z}) \tilde{\rho}_{\mathbf{z}}^h \frac{\delta}{\delta \tilde{\rho}^d_{\mathbf{z}}}\frac{\delta}{\delta \tilde{\rho}^b_{\mathbf{z}}} +2\int_{\mathbf{z}} \tilde{\rho}_{\mathbf{z}}^h \frac{\delta}{\delta \tilde{\rho}^d_{\mathbf{z}}}\int_{\mathbf{y}} \partial_{\mathbf{y}}\tilde{\phi}(\mathbf{x}-\mathbf{y})  \frac{\delta}{\delta \tilde{\rho}^b_{\mathbf{y}}} \partial_{\mathbf{y}}\tilde{\phi}(\mathbf{y} -\mathbf{z})\Bigg).\\
\end{split}
\end{equation}
We have calculated the color structures for the two parts
\begin{equation}
(T^bT^d)_{qa}T^p_{qh}t^p = - (T^pT^bT^d)_{ha} t^p 
\end{equation}
\begin{equation}
(T^{e}T^{d} + T^{d}T^e)_{ha}T^p_{be}t^p = -(T^pT^hT^d)_{ba} t^p +(T^dT^aT^p)_{hb} t^p = (T^pT^hT^a)_{bd}t^p + (T^pT^aT^h)_{bd}t^p
\end{equation}
and performed the integration by parts
\begin{equation}
\begin{split}
&-\int_{\mathbf{z}}\frac{\delta}{\delta \tilde{\rho}^{b}_{\mathbf{z}}} \frac{\delta}{\delta \tilde{\rho}^{d}_{\mathbf{z}}}\int_{\mathbf{y}}\tilde{\phi}(\mathbf{x}-\mathbf{y}) \partial_{\mathbf{y}}^2 \delta(\mathbf{y}-\mathbf{z}) \int_{\mathbf{w}} \tilde{\phi}(\mathbf{y} -\mathbf{w}) \tilde{\rho}^h_{\mathbf{w}}\\
=&-\frac{\delta}{\delta \tilde{\rho}^{b}_{\mathbf{x}}} \frac{\delta}{\delta \tilde{\rho}^{d}_{\mathbf{x}}}  \int_{\mathbf{w}} \tilde{\phi}(\mathbf{x} -\mathbf{w}) \tilde{\rho}^h_{\mathbf{w}} -\int_{\mathbf{z}}\tilde{\phi}(\mathbf{x}-\mathbf{z})\frac{\delta}{\delta \tilde{\rho}^{b}_{\mathbf{z}}} \frac{\delta}{\delta \tilde{\rho}^{d}_{\mathbf{z}}}\tilde{\rho}^h_{\mathbf{z}}\\
&-2\int_{\mathbf{z}}\frac{\delta}{\delta \tilde{\rho}^{b}_{\mathbf{z}}} \frac{\delta}{\delta \tilde{\rho}^{d}_{\mathbf{z}}}\partial_{\mathbf{z}}\tilde{\phi}(\mathbf{x}-\mathbf{z})  \int_{\mathbf{w}} \partial_{\mathbf{z}}\tilde{\phi}(\mathbf{z} -\mathbf{w}) \tilde{\rho}^h_{\mathbf{w}}\\
\end{split}
\end{equation}
and 
\begin{equation}
\begin{split}
&\int_{\mathbf{z}} \tilde{\rho}_{\mathbf{z}}^h \frac{\delta}{\delta \tilde{\rho}^d_{\mathbf{z}}}\int_{\mathbf{y}} \tilde{\phi}(\mathbf{x}-\mathbf{y}) \partial^2_{\mathbf{y}} \frac{\delta}{\delta \tilde{\rho}^b_{\mathbf{y}}} \tilde{\phi}(\mathbf{y} -\mathbf{z})\\
=&\frac{\delta}{\delta \tilde{\rho}^b_{\mathbf{x}}}\int_{\mathbf{z}} \tilde{\phi}(\mathbf{x}-\mathbf{z})\tilde{\rho}_{\mathbf{z}}^h \frac{\delta}{\delta \tilde{\rho}^d_{\mathbf{z}}}  + \int_{\mathbf{z}}\tilde{\phi}(\mathbf{x}-\mathbf{z}) \tilde{\rho}_{\mathbf{z}}^h \frac{\delta}{\delta \tilde{\rho}^d_{\mathbf{z}}}\frac{\delta}{\delta \tilde{\rho}^b_{\mathbf{z}}}  \\
&+2\int_{\mathbf{z}} \tilde{\rho}_{\mathbf{z}}^h \frac{\delta}{\delta \tilde{\rho}^d_{\mathbf{z}}}\int_{\mathbf{y}} \partial_{\mathbf{y}}\tilde{\phi}(\mathbf{x}-\mathbf{y})  \frac{\delta}{\delta \tilde{\rho}^b_{\mathbf{y}}} \partial_{\mathbf{y}}\tilde{\phi}(\mathbf{y} -\mathbf{z}).\\
\end{split}
\end{equation}

The commutation $\int_{\mathbf{z}} [ B_{(1)}^a(\mathbf{z}), A_{(2)}(\mathbf{x}) ]$ , for the relation Eq. \eqref{eq:g3_order_relation} to be correct,  should be equal to 
\begin{equation}
\begin{split}
&-\frac{1}{12} \left( t^aA_{(1)}A_{(2)}+ t^aA_{(2)}A_{(1)} + A_{(1)}A_{(2)}t^a +A_{(2)}A_{(1)}t^a -2A_{(1)}t^aA_{(2)} -2A_{(2)}t^aA_{(1)}\right)\\
=&-\frac{1}{12}\left( \left[\left[ t^a, A_{(1)}\right], A_{(2)}\right] +\left[\left[ t^a, A_{(2)}\right], A_{(1)}\right] \right)\\
=&-\frac{1}{12}\left( [[t^a, t^b], t^m] + [[t^a, t^m], t^b]\right)T^m_{dh}\times  \frac{i}{2} \Bigg(\frac{\delta}{\delta \tilde{\rho}^b_{\mathbf{x}}} \frac{\delta}{\delta\tilde{\rho}^d_{\mathbf{x}}} \int_{\mathbf{w}} \tilde{\phi}(\mathbf{x} -\mathbf{w}) \tilde{\rho}^h_{\mathbf{w}} \\
&+\frac{\delta}{\delta \tilde{\rho}^b_{\mathbf{x}}} \int_{\mathbf{y}}\tilde{\phi}(\mathbf{x}-\mathbf{y}) \frac{\delta}{\delta\tilde{\rho}^d_{\mathbf{y}}}\tilde{\rho}^h_{\mathbf{y}}+2\frac{\delta}{\delta \tilde{\rho}^b_{\mathbf{x}}}\int_{\mathbf{y}} \partial_{\mathbf{y}}\tilde{\phi}(\mathbf{x}-\mathbf{y}) \frac{\delta}{\delta \tilde{\rho}^d_{\mathbf{y}}} \int_{\mathbf{w}} \partial_{\mathbf{y}}\tilde{\phi}(\mathbf{y} -\mathbf{w}) \tilde{\rho}^h_{\mathbf{w}}\Bigg)\\
=&-\frac{1}{12}\left((T^aT^pT^h)_{bd} + (T^pT^aT^h)_{bd}\right)t^p\times  \frac{i}{2} \Bigg(\frac{\delta}{\delta \tilde{\rho}^b_{\mathbf{x}}} \frac{\delta}{\delta\tilde{\rho}^d_{\mathbf{x}}} \int_{\mathbf{w}} \tilde{\phi}(\mathbf{x} -\mathbf{w}) \tilde{\rho}^h_{\mathbf{w}} \\
&+\frac{\delta}{\delta \tilde{\rho}^b_{\mathbf{x}}} \int_{\mathbf{y}}\tilde{\phi}(\mathbf{x}-\mathbf{y}) \frac{\delta}{\delta\tilde{\rho}^d_{\mathbf{y}}}\tilde{\rho}^h_{\mathbf{y}}+2\frac{\delta}{\delta \tilde{\rho}^b_{\mathbf{x}}}\int_{\mathbf{y}} \partial_{\mathbf{y}}\tilde{\phi}(\mathbf{x}-\mathbf{y}) \frac{\delta}{\delta \tilde{\rho}^d_{\mathbf{y}}} \int_{\mathbf{w}} \partial_{\mathbf{y}}\tilde{\phi}(\mathbf{y} -\mathbf{w}) \tilde{\rho}^h_{\mathbf{w}}\Bigg).\\
\end{split}
\end{equation}
We used 
\begin{equation}
\begin{split}
 [[t^a, t^b], t^m] + [[t^a, t^m], t^b]T^m_{dh} =&\left( T^s_{ba}[t^s, t^m] + T^s_{ma}[t^s, t^b]\right)T^m_{dh}\\
 =&(T^s_{ba}T^p_{ms}t^p + T^s_{ma}T^{p}_{bs} t^p)T^m_{dh}\\
 =&\left((T^aT^pT^h)_{bd} + (T^pT^aT^h)_{bd}\right)t^p.
 \end{split}
\end{equation}
The difference is
\begin{equation}
\begin{split}
&\int_{\mathbf{z}} [ B_{(1)}^a(\mathbf{z}), A_{(2)}(\mathbf{x}) ] - \left(-\frac{1}{12}\left( \left[\left[ t^a, A_{(1)}\right], A_{(2)}\right] +\left[\left[ t^a, A_{(2)}\right], A_{(1)}\right] \right)\right)\\
=& \frac{i}{24} \frac{\delta}{\delta \tilde{\rho}^b_{\mathbf{x}}} \frac{\delta}{\delta\tilde{\rho}^d_{\mathbf{x}}} \int_{\mathbf{w}} \tilde{\phi}(\mathbf{x} -\mathbf{w}) \tilde{\rho}^h_{\mathbf{w}}\left((T^aT^pT^h)_{bd} + (T^pT^aT^h)_{bd}+(T^pT^bT^d)_{ha} \right)t^p \\
&+\frac{i}{24}\frac{\delta}{\delta \tilde{\rho}^b_{\mathbf{x}}} \int_{\mathbf{y}}\tilde{\phi}(\mathbf{x}-\mathbf{y}) \frac{\delta}{\delta\tilde{\rho}^d_{\mathbf{y}}}\tilde{\rho}^h_{\mathbf{y}}\left((T^aT^pT^h)_{bd} + 2(T^pT^aT^h)_{bd} + (T^pT^hT^a)_{bd}\right)t^p\\
&+\frac{i}{24} \int_{\mathbf{z}}\tilde{\phi}(\mathbf{x}-\mathbf{z}) \tilde{\rho}_{\mathbf{z}}^h \frac{\delta}{\delta \tilde{\rho}^d_{\mathbf{z}}}\frac{\delta}{\delta \tilde{\rho}^b_{\mathbf{z}}}   ((T^pT^hT^a)_{bd}+ (T^pT^aT^h)_{bd} + (T^pT^bT^d)_{ha}) t^p\\
&+\frac{i}{12}\frac{\delta}{\delta \tilde{\rho}^b_{\mathbf{x}}}\int_{\mathbf{y}} \partial_{\mathbf{y}}\tilde{\phi}(\mathbf{x}-\mathbf{y}) \frac{\delta}{\delta \tilde{\rho}^d_{\mathbf{y}}} \int_{\mathbf{w}} \partial_{\mathbf{y}}\tilde{\phi}(\mathbf{y} -\mathbf{w}) \tilde{\rho}^h_{\mathbf{w}}\left((T^aT^pT^h)_{bd} + (T^pT^aT^h)_{bd}\right)t^p\\
&+\frac{i}{12}\int_{\mathbf{z}}\frac{\delta}{\delta \tilde{\rho}^{b}_{\mathbf{z}}} \frac{\delta}{\delta \tilde{\rho}^{d}_{\mathbf{z}}}\partial_{\mathbf{z}}\tilde{\phi}(\mathbf{x}-\mathbf{z})  \int_{\mathbf{w}} \partial_{\mathbf{z}}\tilde{\phi}(\mathbf{z} -\mathbf{w}) \tilde{\rho}^h_{\mathbf{w}}(T^pT^bT^d)_{ha} t^p \\
&+\frac{i}{12}\int_{\mathbf{z}} \tilde{\rho}_{\mathbf{z}}^h \frac{\delta}{\delta \tilde{\rho}^d_{\mathbf{z}}}\int_{\mathbf{y}} \partial_{\mathbf{y}}\tilde{\phi}(\mathbf{x}-\mathbf{y})  \frac{\delta}{\delta \tilde{\rho}^b_{\mathbf{y}}} \partial_{\mathbf{y}}\tilde{\phi}(\mathbf{y} -\mathbf{z})\left((T^pT^hT^a)_{bd} + (T^pT^aT^h)_{bd}\right)t^p.
\end{split}
\end{equation}
This clearly does not vanish, and so  \eq{commu} is violated at $O(g^3)$.

\end{itemize}

\subsection{ Checking $[Q_L^a, H_{RFT}] = 0$ }
We now directly calculate the commutator of $Q^a_L$ with $H_{RFT}$.

The calculation is organized as expansion in powers of  $g$
\begin{equation}
\begin{split}
\left[Q_L^a, H_{RFT}\right] &= \sum_{n=0}^{\infty}g^n\left[Q_L^a, H_{RFT}\right]_{(n)}  \\
&=\left[Q_L^a, H_{RFT}\right]_{(0)} + g\left[Q_L^a, H_{RFT}\right]_{(1)} + g^2\left[Q_L^a, H_{RFT}\right]_{(2)} + \ldots\\
\end{split}
\end{equation}
where the subscript ``$(n)$'' indicates the $n$-th order in $g$.  

Using   $\left[Q_L^a, V_R \right]=0$ and $\left[Q_L^a, V_L^{\alpha\beta}(\mathbf{x})\right] = - (t^aV_L)^{\alpha\beta} + (V_L t^a)^{\alpha\beta}$, 
we can write
\begin{equation}
\begin{split}\label{comqh}
&\left[Q_L^a , H_{RFT} \right]\\
=& \int_{\mathbf{x}}  \left[Q_L^a, \bar{V}_L^{\beta\gamma}(\mathbf{x})\right]\bar{V}_R^{\delta\alpha}(\mathbf{x})\partial^2_{\mathbf{x}} \left( V_L^{\alpha\beta}(\mathbf{x}) V_R^{\gamma\delta}(\mathbf{x})\right)+ \bar{V}_L^{\beta\gamma}(\mathbf{x}) \left[ Q^a_L, \bar{V}_R^{\delta\alpha}(\mathbf{x})\right]\partial^2_{\mathbf{x}} \left( V_L^{\alpha\beta}(\mathbf{x}) V_R^{\gamma\delta}(\mathbf{x})\right)\\
&+ \left((t^a\bar{V}_L)^{\beta\gamma}\bar{V}_R^{\delta\alpha}(\mathbf{x})- \bar{V}_L^{\beta\gamma}(\mathbf{x}) (\bar{V}_Rt^a)^{\delta\alpha}
\right)\partial^2_{\mathbf{x}} \left(  V_L^{\alpha\beta}(\mathbf{x})V_R^{\gamma\delta}(\mathbf{x})\right)\\
=&\int_{\mathbf{x}}\left\{ \left( \left[Q_L^a, \bar{V}_L^{\beta\gamma}\right]+(t^a\bar{V}_L)^{\beta\gamma}\right)\bar{V}_R^{\delta\alpha}+ \bar{V}_L^{\beta\gamma}\left( \left[ Q^a_L, \bar{V}_R^{\delta\alpha}\right] - (\bar{V}_Rt^a)^{\delta\alpha}\right)\right\}\partial^2_{\mathbf{x}} \left( V_L^{\alpha\beta}(\mathbf{x}) V_R^{\gamma\delta}(\mathbf{x})\right)\\
\end{split}
\end{equation}
Here we have rescaled $H_{RFT}$ by the overall factor $\pi g^2$ for simplicity.

Let us denote 
\begin{equation}
\begin{split}
&d_L^{\beta\gamma} = \left[Q_L^a, \bar{V}_L^{\beta\gamma}\right]+(t^a\bar{V}_L)^{\beta\gamma}\\
&d_R^{\delta\alpha} = \left[ Q^a_L, \bar{V}_R^{\delta\alpha}\right] - (\bar{V}_Rt^a)^{\delta\alpha}\\
\end{split}
\end{equation}
then 
\begin{equation}
\begin{split}
&\left[Q_L^a , H_{RFT} \right]=\int_{\mathbf{x}}\left( d_L^{\beta\gamma}\bar{V}_R^{\delta\alpha}+ \bar{V}_L^{\beta\gamma}d_R^{\delta\alpha}\right)\partial^2_{\mathbf{x}} \left( V_L^{\alpha\beta}(\mathbf{x}) V_R^{\gamma\delta}(\mathbf{x})\right)\\
\end{split}
\end{equation}

Symbolically we write 
\begin{equation}
\begin{split}
&\bar{V}_L  = \mathrm{exp} \left\{ gA_{(1)} + g^2A_{(2)} + g^3A_{(3)} + g^5A_{(5)} + \ldots \right\}\,, \\
&\bar{V}_R  = \mathrm{exp} \left\{ -gA_{(1)} + g^2A_{(2)} - g^3A_{(3)} - g^5A_{(5)} - \ldots \right\}\, ,\\
\end{split}
\end{equation}
and 
\begin{equation}
\begin{split}
&V_L = \mathrm{exp} \left\{gX_{(1)} + g^2X_{(2)} + g^3 X_{(3)} + g^5 X_{(5)}+ \ldots\right\}\, ,\\
&V_R = \mathrm{exp} \left\{-gX_{(1)} + g^2X_{(2)} - g^3 X_{(3)} -g^5 X_{(5)}- \ldots\right\}\, ,\\
\end{split}
\end{equation}
Expansion of $\partial^2_{\mathbf{x}} \left( V_L^{\alpha\beta}(\mathbf{x}) V_R^{\gamma\delta}(\mathbf{x})\right)$ starts at order $g$
\begin{equation}\label{eq:expand_to_second_VLVR}
\begin{split}
&\partial^2_{\mathbf{x}} \left( V_L^{\alpha\beta}(\mathbf{x}) V_R^{\gamma\delta}(\mathbf{x})\right)\\
 = & g \partial^2(-\delta^{\alpha\beta} X_{(1)}^{\gamma\delta} + X_{(1)}^{\alpha\beta}\delta^{\gamma\delta}) \\
&+ g^2\partial^2 \left(\delta^{\alpha\beta} (X_{(2)}^{\gamma\delta} + \frac{1}{2} (X_{(1)}^2)^{\gamma\delta}) + (X_{(2)}^{\alpha\beta}+ \frac{1}{2} (X_{(1)}^2)^{\alpha\beta})\delta^{\gamma\delta} - X_{(1)}^{\alpha\beta}X_{(1)}^{\gamma\delta}\right)\\
&+g^3\partial^2\Bigg(-\delta^{\alpha\beta} \left(X_{(3)}+ \frac{1}{2}(X_{(1)}X_{(2)}+X_{(2)}X_{(1)}) + \frac{1}{3!} X_{(1)}^3\right)^{\gamma\delta}  \\
&\quad + \left(X_{(3)}+ \frac{1}{2}(X_{(1)}X_{(2)}+X_{(2)}X_{(1)}) + \frac{1}{3!} X_{(1)}^3\right)^{\alpha\beta} \delta^{\gamma\delta}\\
&\quad + X_{(1)}^{\alpha\beta}\left(X_{(2)} + \frac{1}{2}X_{(1)}^2\right)^{\gamma\delta} - \left(X_{(2)}+\frac{1}{2} X_{(1)}^2\right)^{\alpha\beta}X_{(1)}^{\gamma\delta}\Bigg)\\
&+\ldots
\end{split}
\end{equation}
Thus the expansion of the Hamiltonian $H_{RFT}$ starts at order $O(g)$. On the other hand
recall that 
\begin{equation}
Q_L^a = \frac{1}{g} \bar{B}_{(-1)} + \bar{B}_{(0)} + g\bar{B}_{(1)} + g^3 \bar{B}_{(3)}+ \ldots
\end{equation}
So the commutation relation $[Q_L^a, H_{RFT}]$ formally starts at order $O(1)$, but from \eq{comqh} it is obvious that at $O(1)$ the commutator vanishes. 

The results of the previous subsection we have calculated explicitly $d_{L(n)}$ for $n\le 3$. Although we have not explicitly calculated $d_{R(n)}$, this calculation up to $n=3$ is identical to that of $d_{L(n)}$ and thus we have
\begin{equation}
\begin{split}
&d_{L (2)} =d_{L(1)}=d_{L(0)} =0\,, \\
& d_{R(2)}=d_{R(1)}=d_{R(0)}=0.\\
&d_{R(3)}=d_{L(3)}
\end{split}
\end{equation}

As a consequence 
\begin{equation}
\begin{split}
&\left[Q_L^a , H_{RFT} \right]_{(1)}=0, \\
& \left[Q_L^a , H_{RFT} \right]_{(2)}=0, \\
&\left[Q_L^a , H_{RFT} \right]_{(3)}=0.\\
\end{split}
\end{equation}
\begin{itemize}
\item  $[Q_L^a, H_{RFT}]_{(4)}$

At order $g^4$, the possible contributions are
\begin{equation}
\left[Q_L^a, H_{RFT}\right]_{(4)} = \int_{\mathbf{x}}\left( d_{L(3)}^{\beta\gamma}\delta^{\delta\alpha}+ \delta^{\beta\gamma}d_{R(3)}^{\delta\alpha}\right)\partial^2_{\mathbf{x}} \left( V_L^{\alpha\beta}(\mathbf{x}) V_R^{\gamma\delta}(\mathbf{x})\right)_{(1)}\\
\end{equation}
However, this expression vanishes after substituting the first order result in Eq. \eqref{eq:expand_to_second_VLVR} and using the identity
\begin{equation}\label{eq:1st_cancelling}
(-\delta^{\alpha\beta} X_{(1)}^{\gamma\delta} + X_{(1)}^{\alpha\beta}\delta^{\gamma\delta}) \delta^{\delta\alpha} = 0. 
\end{equation}
Therefore
\begin{equation}
\left[Q_L^a, H_{RFT}\right]_{(4)} =0
\end{equation}

\item $[Q_L^a, H_{RFT}]_{(5)}$

The possible contributions at  order $g^5$ are
\begin{equation}
\begin{split}
&\left[Q_L^a, H_{RFT}\right]_{(5)}\\
= & \int_{\mathbf{x}} \left( d_{L(3)}^{\beta\gamma}\delta^{\delta\alpha}+ \delta^{\beta\gamma}d_{R(3)}^{\delta\alpha}\right)\partial^2_{\mathbf{x}} \left( V_L^{\alpha\beta} V_R^{\gamma\delta}\right)_{(2)}\\
&+\int_{\mathbf{x}}\left( d_{L(3)}^{\beta\gamma} (-A_{(1)}^{\delta\alpha})+A_{(1)}^{\beta\gamma}d_{R(3)}^{\delta\alpha}\right)\partial^2_{\mathbf{x}} \left( V_L^{\alpha\beta} V_R^{\gamma\delta}\right)_{(1)}\\
&+ \int_{\mathbf{x}}\left( d_{L(4)}^{\beta\gamma}\delta^{\delta\alpha}+ \delta^{\beta\gamma}d_{R(4)}^{\delta\alpha}\right)\partial^2_{\mathbf{x}} \left( V_L^{\alpha\beta} V_R^{\gamma\delta}\right)_{(1)}\\
\end{split}
\end{equation}
Note that the last term vanishes due to Eq. \eqref{eq:1st_cancelling}. Let us focus on the other two terms. From Eq. \eqref{eq:expand_to_second_VLVR}, one obtains
\begin{equation}\label{eq:02}
\begin{split}
&  \int_{\mathbf{x}} \left( d_{L(3)}^{\beta\gamma}\delta^{\delta\alpha}+ \delta^{\beta\gamma}d_{R(3)}^{\delta\alpha}\right)\partial^2_{\mathbf{x}} \left( V_L^{\alpha\beta} V_R^{\gamma\delta}\right)_{(2)}\\
=&  \int_{\mathbf{x}}  2\mathrm{Tr} \left[(d_{L(3)}  + d_{R(3)}) \partial^2 X_{(2)}\right]
\end{split}
\end{equation}
and
\begin{equation}\label{eq:11}
\begin{split}
&\int_{\mathbf{x}}\left( d_{L(3)}^{\beta\gamma}(- A_{(1)}^{\delta\alpha})+(A_{(1)}^{\beta\gamma})d_{R(3)}^{\delta\alpha}\right)\partial^2_{\mathbf{x}} \left( V_L^{\alpha\beta} V_R^{\gamma\delta}\right)_{(1)}\\
=& \int_{\mathbf{x}}\mathrm{Tr}\left[(d_{L(3)}+d_{R(3)})(\partial^2X_{(1)}A_{(1)} - A_{(1)}\partial^2 X_{(1)})\right]
\end{split}
\end{equation}
Recall the expressions
\begin{equation}
\begin{split}
&A_{(1)} (\mathbf{x})= t^a\frac{\delta}{\delta \tilde{\rho}^a(\mathbf{x})}, \quad  X_{(1)}(\mathbf{x}) = i\int_{\mathbf{y}}\phi(\mathbf{x}-\mathbf{y}) t^a \tilde{\rho}^a(\mathbf{y}),\\
&X_{(2)} = i\int_{\mathbf{y}}\phi(\mathbf{x}-\mathbf{y}) t^a \left(-\frac{1}{2} \frac{\delta}{\delta\tilde{\rho}^e}T^e_{ba}\rho^b\right).\\
\end{split}
\end{equation}
Using this one obtains
\begin{equation}\label{eq:11and02_cancellation}
\begin{split}
\partial^2X_{(1)}A_{(1)} - A_{(1)}\partial^2 X_{(1)} =& i(t^bt^e-t^et^b)\tilde{\rho}^b(\mathbf{x})  \frac{\delta}{\delta\tilde{\rho}^e(\mathbf{x})} = i T^a_{eb} t^a \tilde{\rho}^b(\mathbf{x})  \frac{\delta}{\delta\tilde{\rho}^e(\mathbf{x})} \\
= &-2\partial^2 X_{(2)}\, .\\
\end{split}
\end{equation}
As a consequence the contributions in Eq. \eqref{eq:02} and Eq. \eqref{eq:11} cancel each other. 

We have  proved that $[Q_L^a, H_{RFT}]_{(5)}=0.$

\item $[Q_L^a, H_{RFT}]_{(6)}$,

The possible contributions at  order $g^6$ are
\begin{equation}
\begin{split}
&\left[Q_L^a, H_{RFT}\right]_{(6)}\\
=&\int_{\mathbf{x}}\left( d_{L(5)}^{\beta\gamma}\delta^{\delta\alpha}+ \delta^{\beta\gamma}d_{R(5)}^{\delta\alpha}\right)\partial^2_{\mathbf{x}} \left( V_L^{\alpha\beta} V_R^{\gamma\delta}\right)_{(1)}\\
&+ \int_{\mathbf{x}}\left( d_{L(4)}^{\beta\gamma} \bar{V}_{R(1)}^{\delta\alpha}+\bar{V}_{L(1)}^{\beta\gamma}d_{R(4)}^{\delta\alpha}\right)\partial^2_{\mathbf{x}} \left( V_L^{\alpha\beta} V_R^{\gamma\delta}\right)_{(1)}\\
& + \int_{\mathbf{x}} \left( d_{L(4)}^{\beta\gamma}\delta^{\delta\alpha}+ \delta^{\beta\gamma}d_{R(4)}^{\delta\alpha}\right)\partial^2_{\mathbf{x}} \left( V_L^{\alpha\beta} V_R^{\gamma\delta}\right)_{(2)}\\
&+ \int_{\mathbf{x}}\left( d_{L(3)}^{\beta\gamma} \bar{V}_{R(2)}^{\delta\alpha}+\bar{V}_{L(2)}^{\beta\gamma}d_{R(3)}^{\delta\alpha}\right)\partial^2_{\mathbf{x}} \left( V_L^{\alpha\beta} V_R^{\gamma\delta}\right)_{(1)}\\
&+ \int_{\mathbf{x}}\left( d_{L(3)}^{\beta\gamma} \bar{V}_{R(1)}^{\delta\alpha}+\bar{V}_{L(1)}^{\beta\gamma}d_{R(3)}^{\delta\alpha}\right)\partial^2_{\mathbf{x}} \left( V_L^{\alpha\beta} V_R^{\gamma\delta}\right)_{(2)}\\
&+ \int_{\mathbf{x}}\left( d_{L(3)}^{\beta\gamma} \delta^{\delta\alpha}+\delta^{\beta\gamma}d_{R(3)}^{\delta\alpha}\right)\partial^2_{\mathbf{x}} \left( V_L^{\alpha\beta} V_R^{\gamma\delta}\right)_{(3)}\\
\end{split}
\end{equation}
The first term vanishes due to Eq. \eqref{eq:1st_cancelling}. The second and third terms add up to zero because of Eq. \eqref{eq:11and02_cancellation}.  Now we focus on the fourth, fifth and sixth terms.

For the sixth term, note that
\begin{equation}
\begin{split}
&\delta^{\delta\alpha} \left( V_L^{\alpha\beta} V_R^{\gamma\delta}\right)_{(3)} = (X_{(2)}X_{(1)} - X_{(1)}X_{(2)})^{\gamma\beta}\\
&\delta^{\beta\gamma} \left( V_L^{\alpha\beta} V_R^{\gamma\delta}\right)_{(3)} =  \left(X_{(1)}X_{(2)}-X_{(2)}X_{(1)}\right)^{\alpha\delta}
\end{split}
\end{equation}
Using these relations, one calculates 
\begin{equation}
\begin{split}
& \int_{\mathbf{x}}\left( d_{L(3)}^{\beta\gamma} \delta^{\delta\alpha}+\delta^{\beta\gamma}d_{R(3)}^{\delta\alpha}\right)\partial^2_{\mathbf{x}} \left( V_L^{\alpha\beta} V_R^{\gamma\delta}\right)_{(3)}\\
=& \int_{\mathbf{x}}\mathrm{Tr}\left[ (d_{L(3)}-d_{R(3)})\partial^2_{\mathbf{x}} (X_{(2)}X_{(1)} - X_{(1)}X_{(2)})\right] \\
=&0\\
\end{split}
\end{equation}
Here we have used the relation $d_{L(3)} = d_{R(3)}$. 

For the fifth term, 
\begin{equation}
\begin{split}
d_{L(3)}^{\beta\gamma}\bar{V}^{\delta\alpha}_{R(1)} \partial^2(V_L^{\alpha\beta}V_R^{\gamma\delta})_{(2)}  =& -\mathrm{Tr}\left[d_{L(3)} \left(X_{(2)}+\frac{1}{2}X_{(1)}^2\right)A_{(1)}\right] - \mathrm{Tr}\left[d_{L(3)}A_{(1)} \left(X_{(2)}+\frac{1}{2}X_{(1)}^2\right)\right] \\
&+ \mathrm{Tr}\left[d_{L(3)}X_{(1)}A_{(1)}X_{(1)}\right] \\
\end{split}
\end{equation}
and 
\begin{equation}
\begin{split}
\bar{V}^{\beta\gamma}_{(1)}d_{R(3)}^{\delta\alpha} (V_L^{\alpha\beta} V_R^{\gamma\delta})_{(2)} =&\mathrm{Tr} \left[d_{R(3)}A_{(1)}\left(X_{(2)}+\frac{1}{2}X_{(1)}^2\right)\right] + \mathrm{Tr}\left[d_{R(3)}\left(X_{(2)}+\frac{1}{2}X_{(1)}^2\right)A_{(1)}\right]\\
& - \mathrm{Tr} \left[d_{R(3)} X_{(1)}A_{(1)}X_{(1)}\right] \\
\end{split}
\end{equation}
The sum of these two terms vanishes due to equality $d_{L(3)} =d_{R(3)}$. 

For the fourth term
\begin{equation}
d_{L(3)}^{\beta\gamma} \bar{V}_{R(2)}^{\delta\alpha}(V_L^{\alpha\beta}V_R^{\gamma\delta})_{(1)} = -\mathrm{Tr} \left[d_{L(3)}X_{(1)}(A_{(2)}+\frac{1}{2}A_{(1)}^2)\right] + \mathrm{Tr} \left[d_{L(3)}(A_{(2)}+\frac{1}{2}A_{(1)}^2) X_{(1)} \right]
\end{equation} 
and 
\begin{equation}
\bar{V}_{L(2)}^{\beta\gamma} d_{R(3)}^{\delta\alpha} (V_L^{\alpha\beta}V_R^{\gamma\delta})_{(1)} = -\mathrm{Tr}\left[d_{R(3)}(A_{(2)}+\frac{1}{2}A_{(1)}^2)X_{(1)}\right] + \mathrm{Tr}\left[d_{R(3)}X_{(1)}(A_{(2)}+\frac{1}{2}A_{(1)}^2)\right]
\end{equation}
These two terms  also cancel  each other due to $d_{L(3)} =d_{R(3)}$. 
We therefore proved that 
$
[Q_L^a, H_{RFT}]_{(6)} =0.
$

\end{itemize}

Thus we see that up to order $O(g^6)$ the left rotation generator $Q^a_L$ commutes with the Hamiltonian.


\newpage

\begin{thebibliography}{99}
 
 \bibitem{gribov} V.~N.~Gribov,
  Sov.\ Phys.\ JETP {\bf 26}, 414 (1968)
  [Zh.\ Eksp.\ Teor.\ Fiz.\  {\bf 53}, 654 (1967)].


\bibitem{BFKL}
 E. A. Kuraev, L. N. Lipatov, and F. S. Fadin, { Sov. Phys.
JETP}
                {\bf 45}, 199 (1977); \,\,\, \\
Ya. Ya. Balitsky and L. N. Lipatov,
               { Sov. J. Nucl. Phys.}\, {\bf 28}, 22 (1978).
               
               
\bibitem{glr} L. Gribov, E. Levin and M. Ryskin, Phys. Rept. {\bf 100}, 1, 1983.

\bibitem{MUPA}
A. H. Mueller and J. Qiu, 
Nucl. Phys.  B  {\bf 268} (1986) 427.
\bibitem{MUDI}
  A.~H.~Mueller,
  Nucl.\ Phys.\ B {\bf 415} (1994) 373;\,\,\,
  Nucl.\ Phys.\ B {\bf 437} (1995) 107;\\
      A.~H.~Mueller and B.~Patel, Nucl. Phys. B {\bf 425}, 471, 1994.         
               
   \bibitem{LIREV}
                L.~N.~Lipatov,
  Phys.\ Rept.\  {\bf 286} (1997) 131.
               
\bibitem{LipatovFT}
  L.~N.~Lipatov,
  Nucl.\ Phys.\ B {\bf 365}, 614 (1991),
  Nucl.\ Phys.\ B {\bf 452}, 369 (1995), \\
R.~Kirschner, L.~N.~Lipatov and L.~Szymanowski,
  Nucl.\ Phys.\ B {\bf 425}, 579 (1994),
  Phys.\ Rev.\ D {\bf 51}, 838 (1995).


  \bibitem{bartels} J.~Bartels, Z.Phys. C {\bf 60}, 471 (1993);\\
J.~Bartels and M.~Wusthoff, Z. Phys. C {\bf 66}, 157 (1995);\,\,\,
 J.~Bartels and C.~Ewerz, JHEP  {\bf 9909}, 026 (1999), \\
   C.~Ewerz,
  JHEP {\bf 0104} (2001) 031.

\bibitem{BKP}
J.~Bartels,
%
Nucl.\ Phys.\  B {\bf 175}, 365 (1980);\\
J.~Kwiecinski and M.~Praszalowicz,
%
Phys.\ Lett.\ B   {\bf 94}, 413 (1980).


 
 \bibitem{KLLL} 
 A.~Kovner, E.~Levin, M.~Li and M.~Lublinsky,
{\it``The JIMWLK evolution and the s-channel unitarity,''}
[arXiv:2006.15126 [hep-ph]].
 
  \bibitem{KLL} A. Kovner, E.Levin and M. Lublinsky, JHEP {\bf 1608}   (2016) 031.
  


  
  
  \bibitem{mv} L.~McLerran and R.~Venugopalan, Phys. Rev. D {\bf 49},  2233-2241, (1994);  Phys. Rev. D {\bf 49}, 3352-3355, (1994). 

\bibitem{Salam}
A.~H.~Mueller and G.~P.~Salam,
  Nucl.\ Phys.\ B {\bf 475}, 293 (1996);\\
  G.~P.~Salam,
  Nucl.\ Phys.\ B {\bf 461}, 512 (1996).
  
   \bibitem{KOLE}
  Y.~V.~Kovchegov and E.~Levin,
  Nucl.\ Phys.\ B {\bf 577} (2000) 221.

  
  
  
   \bibitem{BRN}
M. A. Braun,
Eur. Phys. J.  C {\bf 16} (2000) 337;\\
M. A. Braun and G. P. Vacca,
Eur. Phys. J.  C {\bf 6} (1999) 147; \\
J.~Bartels, M.~Braun and G.~P.~Vacca,
  Eur.\ Phys.\ J.\ C {\bf 40}, 419 (2005);
\\
 J.~Bartels, L.~N.~Lipatov and G.~P.~Vacca,
  Nucl.\ Phys.\ B {\bf 706}, 391 (2005).

\bibitem{braun}  M.~A.~Braun,
  Phys.\ Lett.\ B {\bf 483}, 115 (2000), 
  Eur.\ Phys.\ J.\ C {\bf 33}, 113 (2004);
  Phys.\ Lett.\ B {\bf 632}, 297 (2006).  
  
 
\bibitem{BK}
I.~Balitsky,
{Phys.\ Rev.}  D  {\bf 60}, 014020 (1999);
Y.~V.~Kovchegov,
{Phys.\ Rev.}  D  {\bf 60}, 034008  (1999).




\bibitem{AKLL}
  T.~Altinoluk, C.~Contreras, A.~Kovner, E.~Levin, M.~Lublinsky and A.~Shulkim,
  Int.\ J.\ Mod.\ Phys.\ Conf.\ Ser.\  {\bf 25} (2014) 1460025;\,\,\,\\
   T.~Altinoluk, N.~Armesto, A.~Kovner, E.~Levin and M.~Lublinsky,
  JHEP {\bf 1408} (2014) 007.
  \bibitem{AKLL1}
 T.~Altinoluk, A.~Kovner, E.~Levin and M.~Lublinsky,
  JHEP {\bf 1404} (2014) 075;
  T.~Altinoluk, C.~Contreras, A.~Kovner, E.~Levin, M.~Lublinsky and A.~Shulkin,
  JHEP {\bf 1309} (2013) 115.

         
 
\bibitem{jimwlk} J. Jalilian Marian, A. Kovner, A. Leonidov and H. Weigert,
{Nucl. Phys.} B {\bf  504} 415 (1997);  
{ Phys. Rev.}   D {\bf 59} 014014 (1999);  \\
J. Jalilian Marian, A. Kovner and H. Weigert, Phys. Rev. D  {\bf 59} 
014015 (1999); \\
A. Kovner and J.G. Milhano, {Phys. Rev.} D  {\bf 61} 014012 (2000); \\
 A. Kovner, J.G. Milhano and H. Weigert,
{ Phys. Rev.}  D {\bf 62} 114005 (2000); \\
 H. Weigert, { Nucl.Phys.} A  {\bf  703} (2002) 823.
 \bibitem{klwmij} A. Kovner and M. Lublinsky; Phys. Rev.   D {\bf 71},  085004 (2005).
 
 \bibitem{cgc}  E.Iancu, A. Leonidov and L. McLerran, {Nucl. Phys.}  A
{\bf  692} (2001) 583; {Phys. Lett.}  B {\bf 
510} (2001) 133;\\
E. Ferreiro, E. Iancu, A. Leonidov, L. McLerran;  
{Nucl. Phys.} A {\bf 703} (2002) 489.

 \bibitem{reggeon} A. Kovner and M. Lublinsky, JHEP {\bf 0702}, 058 (2007).

  
  \bibitem{nlo} I. Balitsky and G. Chirilli,  Nucl.Phys.B {\bf 82} 2 (2009) 45-87; Phys.Rev.D {\bf 88} (2013) 111501; \\ 
  A. Kovner, M. Lublinsky and Y. Mulian, Phys.Rev.D {\bf  89} (2014) 6, 061704; JHEP {\bf 08} (2014) 114;\,\,\\  M. Lublinsky and Y. Mulian, JHEP {\bf 05} (2017) 097.
  
  \bibitem{caron} S. Caron-Huot, JHEP {\bf 03} (2018) 036;\,\,\, S. Caron-Huot and M. Herranen, JHEP {\bf 02} (2018) 058 .


\bibitem{pomloops}
  A.~H.~Mueller and A.~I.~Shoshi,
  Nucl.\ Phys.\ B {\bf 692}, 175 (2004);\,\,\,
  E. Iancu adn A. Mueller, Nucl.Phys.A  {\bf 730} (2004) 494-513 ;
  E.~Iancu and D.~N.~Triantafyllopoulos,
  Nucl.\ Phys.\ A {\bf 756} (2005) 419;
  Phys.\ Lett.\ B {\bf 610} (2005) 253; \,\,\,
    A.~H.~Mueller, A.~I.~Shoshi and S.~M.~H.~Wong,
  Nucl.\ Phys.\ B {\bf 715} (2005) 440;\,\,\,
   E.~Levin and M.~Lublinsky,
  Nucl.\ Phys.\ A {\bf 763} (2005) 172; \,\,\,
E. Levin, J. Miller and A. Prygarin,
  Nucl.\ Phys.\  {\bf A806 } (2008)  245;\,\,\,
   E.~Iancu, G.~Soyez and D.~N.~Triantafyllopoulos,
  Nucl.\ Phys.\ A {\bf 768} (2006) 194;
T.~Altinoluk, A.~Kovner, E.~Levin and M.~Lublinsky,
JHEP \textbf{04}, 075 (2014).


\bibitem{KLduality} 
  A.~Kovner and M.~Lublinsky,
  Phys.\ Rev.\ Lett.\  {\bf 94}, 181603 (2005).
  
   
 
\bibitem{diamond} 
Y. Hatta, E. Iancu, L. McLerran, A. Stasto and D.N. Triantafyllopoulos , Nucl.Phys.  A  {\bf 764 }(2006) 423.
  
  
  
  
\bibitem{Balitsky05}
  I.~Balitsky,
  Phys.\ Rev.\ D {\bf 72}, 074027 (2005).
  
   \bibitem{ddd} A. Kovner and M. Lublinsky,  
  Phys.Rev. D {\bf 72} (2005) 074023.

 \bibitem{likovner} M. Li and A. Kovner, JHEP  {\bf  05} (2020) 036.  
  
  \bibitem{motyka} S. Bondarenko and L. Motyka,  Phys. Rev. D {\bf 75} (2007) 114015.
  
  \bibitem{yin} 
A. Kovner and M. Lublinsky, Nucl.Phys. A {\bf 779} (2006) 220-243.  

    
  \bibitem{KLremark} 
  A.~Kovner and M.~Lublinsky,
JHEP {\bf 0503}, 001 (2005).


  \bibitem{KLremark2} 
  A.~Kovner and M.~Lublinsky,
  Nucl.\ Phys.\ A {\bf 767} 171 (2006).
  
   \bibitem{Kancheli} O.V. Kancheli, 	arXiv:2003.04654 [hep-ph]. 

    
\end{thebibliography}
\end{document}